\documentclass[12pt,a4paper]{article}
\usepackage{amsmath}
\usepackage{graphicx}
\usepackage{here}
\usepackage{tikz}
\usepackage{bm}
\usetikzlibrary{matrix}
\usepackage{overcite} 
\usepackage{dcolumn} 
\newcolumntype{.}{D{.}{.}{1}}

\begin{document}
\newcommand{\rhor}{\ensuremath{\rho(\mathbf{r})}}
\newcommand{\mr}{\ensuremath{(\mathbf{r})}}
\newcommand{\dr}{\ensuremath{{\rm d}}}
\newcommand\Item[1][]{%
  \ifx\relax#1\relax  \item \else \item[#1] \fi
  \abovedisplayskip=0pt\abovedisplayshortskip=0pt~\vspace*{-\baselineskip}}
\begin{center}

\vspace*{1cm}

{\LARGE\bf
Tensor Network States\\[1.5ex] with Three-Site Correlators
}

\vspace{1cm}

{\large Arseny Kovyrshin and Markus Reiher\footnote{Corresponding author: markus.reiher@phys.chem.ethz.ch}}
\\[2ex]
ETH Z\"urich, Laboratorium f\"ur Physikalische Chemie, 
Vladimir-Prelog-Weg 2, \\
8093 Z\"urich, Switzerland \\[2ex]

\begin{abstract}
We present a detailed analysis of various tensor network parameterizations within the Complete Graph Tensor Network States (CGTNS) approach. 
We extend our 2-site CGTNS scheme by introducing 3-site correlators. For this we devise three different strategies. The first relies solely 
on 3-site correlators and the second on 3-site correlators added on top of the 2-site correlator ansatz. To avoid an inflation 
of the variational space introduced by higher-order correlators, we limit the number of higher-order correlators to the 
most significant ones in the third strategy. Approaches for the selection of these most significant correlators are discussed. The sextet and doublet spin states of 
the spin-crossover complex manganocene serve as a numerical test case. In general, the CGTNS scheme achieves a remarkable accuracy for a 
significantly reduced size of the variational space. The advantages, drawbacks, and limitations of all CGTNS parameterizations investigated 
are rigorously discussed. 
\end{abstract}

\vfill

\end{center}

\begin{tabbing}
Date:   \quad\= May 22, 2016 \\
\end{tabbing}

\newpage
\section{Introduction}
Electronic structure theory aims at providing accurate properties of molecules in 
their electronic ground and excited states. However, systematically improvable wave function methods are 
often computationally expensive and can even become unfeasible. This dilemma is particularly pressing for systems with strong 
static electron correlation, i.e., for those with dense one-electron states around the Fermi energy level showing a small gap.

For such cases, the Density-Matrix Renormalization Group (DMRG) method \cite{whit1992,scho2005,scho2011} has evolved as a powerful 
alternative to exact diagonalization techniques such as Complete Active Space Self-Consistent Field (CAS-SCF) 
approaches\cite{lege2008,chan2008a,chan2009a,reih2010,chan2011,marti2011,kell2014,kura2014,wout2014a,yana2015,szal2015,knec2016}. 
The success of DMRG is due to a polynomial scaling of the computational cost with respect to an increasing number of active orbitals, in contrast 
to the exponential scaling\cite{molcas8} of CAS-based approaches. It was shown by \"Ostlund and Rommer\cite{ostl1995} 
that the DMRG optimizes Matrix Product States (MPS) --- one-dimensional chains of tensors 
that are a consequence of the algorithm which imposes a one-dimensional ordering of the 
molecular orbitals in the construction process of the total basis states. 

In contrast to one-dimensional spin chains in solid-state physics, molecular systems governed by the full Coulomb 
interaction in general feature multidimensional entanglement for which the linear MPS ansatz is not well suited. 
This in turn may lead to convergence problems. Still, the DMRG optimization of MPSs can be beneficial for strongly 
correlated molecules if other approaches are unfeasible as we pointed out for transition metal complexes\cite{mart2008}. 
Moreover, dynamic correlation effects have to be considered --- either a posteriori by perturbation 
theory\cite{kura2011,shar2014,knec2016,shen2015} or from the outset by, for example, short--range DFT\cite{jensen,hede2015}.

One can overcome the problem of complex, multidimensional entanglement patterns by generalizing the MPS ansatz (see Ref.\ \citenum{szal2015}
for a discussion in the context of electronic structure theory). 
In the field of solid-state physics, the Tensor Product Variational Approach 
(TPVA)\cite{nish2000,nish2001,gend2002,gend2003}, String Bond States (SBS)\cite{schu2008}, Projected Entangled 
Pair States (PEPS)\cite{vers2004a}, Multiscale Entanglement Renormalization Ansatz (MERA)\cite{even2014}, Entangled 
Plaquette States\cite{mezz2009} (EPS), and Correlator Product States (CPS)\cite{chan2009} attempt to generalize the 
MPS ansatz and describe multidimensional entanglement. These approaches constitute a new family of states called 
Tensor Network States (TNS).

Complete-Graph Tensor Network States (CGTNS)\cite{mart2010b} were the first TNS application in quantum 
chemistry employing the full electronic Hamiltonian. The complexity of the high-dimensional coefficient 
tensor was reduced by breaking it down into a complete-graph tensor network, in which all spin orbitals 
are connected with each other by 2-site correlators. The number of variational parameters is explicitly 
defined by the number of spin orbital pairs, which limits the variational freedom of the CGTNS ansatz for 
large active spaces. 

Another TNS approach explored in the field of quantum chemistry are Tree Tensor Network States 
(TTNSs)\cite{murg2010,barc2011}. While representing an interesting class of quantum states, the 
optimization of TTNS parameters can be cumbersome and non-competitive when compared to efficient traditional 
quantum chemical methods. In 2013, Nakatani and Chan proposed a TTNS variant for the 
full electronic Hamiltonian\cite{naka2013} that overcomes this problem. Tensors are connected as defined 
by a tree graph in TTNS, which attempt to map the molecular structure. Due to the absence of loops in their TTNS 
ansatz, Nakatani and Chan could apply the DMRG optimization algorithm. 

The MPS ansatz is computationally very efficient, but truly reliable only for encoding a sequential entanglement 
structure. The TTNS scheme provides a more general description of entanglement which is, in the Nakatani--Chan formulation, 
of similar computational efficiency as MPS-DMRG, but it still imposes restrictions on the entanglement structure. 
CGTNS, in principle, does not restrict the entanglement pattern and works as well for multidimensional 
entanglement as it does for one-dimensional entanglement. However, the optimization of 2-site correlators is difficult as an 
efficient global and local optimization strategy is required. It is desirable to have an ansatz which adjusts the 
number of variational parameters to the system under study. Therefore, here we propose the concept of a CGTNS ansatz
that starts from 2-site correlators and gradually include higher-order correlators. We then explore different 
optimization strategies to assess its potential for actual applications in molecular physics and chemistry.

\section{Theory}
\subsection{Exact Solution}
The eigenstate $\left| \Psi \right\rangle$ of electronic Hamiltonian $H$ for an $N$-electron molecular system in 
non-relativistic quantum mechanics can be expressed as a linear combination of all Slater determinants contained 
in the $M\choose N$-dimensional subspace $F(M,N)$ of the $2^M$-dimensional Fock space $F(M)$, 
\begin{equation}\label{eq:gen_state}
\left| \Psi \right\rangle =\sum_{n_1 n_2 \ldots n_M} C_{n_1 n_2 \ldots n_M} \left| n_1 n_2 \ldots n_M \right\rangle,
\end{equation}
where $M$ is the total number of spin orbitals constructed within a given one-electron basis set, and $\left| n_1 n_2 \ldots n_M \right\rangle$
is an occupation number vector (ONV) representing a Slater determinant in the second quantization formalism.  It is 
constructed as a tensor product of spin orbitals (sites) $\left| n_i \right\rangle$,
\begin{equation}
\left| n_1 n_2 \ldots n_M \right\rangle \equiv \left| n_1 \right\rangle \otimes \left| n_2 \right\rangle \otimes \ldots \otimes  \left| n_M \right\rangle.
\end{equation}
Every spin orbital can be ``unoccupied'' or ``occupied'', $n_i=\{0,1\}$, which yields a dimension of the local 
Hilbert space of two. Eq.\ (\ref{eq:gen_state}) is called a Full Configuration Interaction (FCI) expansion. The 
coefficients $C_{n_1 n_2 \ldots n_M}$ in Eq.~(\ref{eq:gen_state}) are obtained according to the variational principle which 
leads to the eigenvalue problem 
\begin{equation}
\mathbf{HC}=E\mathbf{C},
\end{equation}
where $\mathbf{H}$ is the matrix representation of $H$ in the determinant basis and $\mathbf{C}$ is a vector of 
$C_{n_1 n_2 \ldots n_M}$ coefficients for an electronic state of energy $E$. Since the number of Slater determinants 
that can be generated by distributing $N$ electrons among $M$ spin orbitals scales exponentially\cite{molcas8}, the 
$C_{n_1 n_2 \ldots n_M}$ can be found only for systems of limited size of up to about 18 electrons in 18 spatial orbitals. 
Hence, approximations are desirable for approaching the FCI solution in a given orbital space with more orbitals and electrons.

\subsection{Tensor Network Decomposition}
Tensor network approximations were suggested to reduce the dimensionality of a strongly correlated system. The TNS 
parameterizations mentioned in the introduction were investigated for model Hamiltonians such as the Heisenberg, Hubbard, 
Potts, and Ising Hamiltonians. The idea behind parameterizations such as TPVA, SBS, EPS, and CPS is similar, which 
can be seen in the work by Changlani {\it et al.}\cite{chan2009} and other CPS studies\cite{neus2012,neus2012a}.
They afford a factorization of the high-dimensional coefficient tensor $C_{n_1 n_2 \ldots n_M}$ into a product of nearest-neighbor 
2-site correlator elements $C_{n_i n_j}^{[ij]}$,
\begin{equation}\label{eq:CPS}
\left| \Psi_{\rm CPS} \right\rangle =\sum_{n_1 n_2 \ldots n_M} \prod_{\langle i j \rangle} 
C_{n_i n_j}^{[ij]} \left| n_1 n_2 \ldots n_M \right\rangle,
\end{equation}
where $\langle i j \rangle$ indicates that only neighboring sites are taken into account. The correlators are represented by the 
second order tensors $\mathbf{C}^{[ij]}$,
\begin{equation}\label{eq:tensor2}
\mathbf{C}^{[ij]} \equiv \left[ 
\begin{array}{cc}
C_{00}^{[ij]} & C_{01}^{[ij]} \\
C_{10}^{[ij]} & C_{11}^{[ij]}
\end{array}
\right],
\end{equation}
for each pair of neighboring spin orbitals $i$ and $j$.

\subsubsection{2-site Correlator Ansatz}\label{sec:2site}
We extended\cite{mart2010b} the nearest-neighbors ansatz to {\it all} possible 2-site correlators,
\begin{equation}\label{eq:CGTNS}
\left| \Psi^{2s}_{\rm CGTNS} \right\rangle =\sum_{n_1 n_2 \ldots n_M} \prod_{i \le j} C_{n_i n_j}^{[ij]} 
\left| n_1 n_2 \ldots n_M \right\rangle,
\end{equation}
which we therefore denoted Complete Graph Tensor Network States (CGTNS). Note that the notation for the CGTNS ansatz in Eq.\ (\ref{eq:CGTNS}) 
is different from the one presented in the original paper\cite{mart2010b} in order to better discriminate the different tensor networks 
explored in this work. The total number of correlators in this ansatz is equal to $1/2M(M+1)$. Taking into account that each 
correlator is represented by a tensor of second order, Eq.~(\ref{eq:tensor2}) with $q^2$ elements ($q=2$ for spin orbitals 
and $q=4$ for spatial orbitals), the total number of variational parameters is equal to $1/2M(M+1)q^2$. 

It is possible to avoid correlator matrices in CGTNS corresponding to interactions of certain sites with themselves --- which we may call 
self-interaction ($si$) correlators -- $\mathbf{C}^{[ii]}$ --- and to obtain the following ansatz
\begin{equation}\label{eq:CGTNS}
\mid \Psi^{2s/si}_{\rm CGTNS} \rangle =\sum_{n_1 n_2 \ldots n_M} \prod_{i < j} C_{n_i n_j}^{[ij]}
\left| n_1 n_2 \ldots n_M \right\rangle.
\end{equation}
This removes $M$ $si$ correlators from the ansatz [$1/2M(M-1)$ correlators] without serious loss in accuracy as we shall 
demonstrate in Section \ref{sec:sextet}. The graphical representation of such a tensor network ansatz at the  example 
of a four-site system is shown in Figure \ref{fig:graphs} a). 

Note that the CGTNS ansatz is related not only to TPVA, SBS, EPS, and CPS, but also 
to the Antisymmetric Products of Nonorthogonal Geminals ansatz\cite{john2013,lima2013}. The latter 
may be considered as a special case of $\left| \Psi^{2s}_{\rm CGTNS} \right\rangle$ where only the correlators 
between spin orbitals with the same spatial part are employed.

\subsubsection{3-site Correlator Ansatz}
\begin{figure}[H]
\caption{Graphical representations of the a) $\Psi^{2s/si}_{\rm CGTNS}$ and b)$\Psi^{3s/si}_{\rm CGTNS}$ ans\"atze for a system 
containing four sites. The blue vertices represent sites, while the connecting black lines represent correlators.\label{fig:graphs}}
\begin{center}
\includegraphics[scale=0.3]{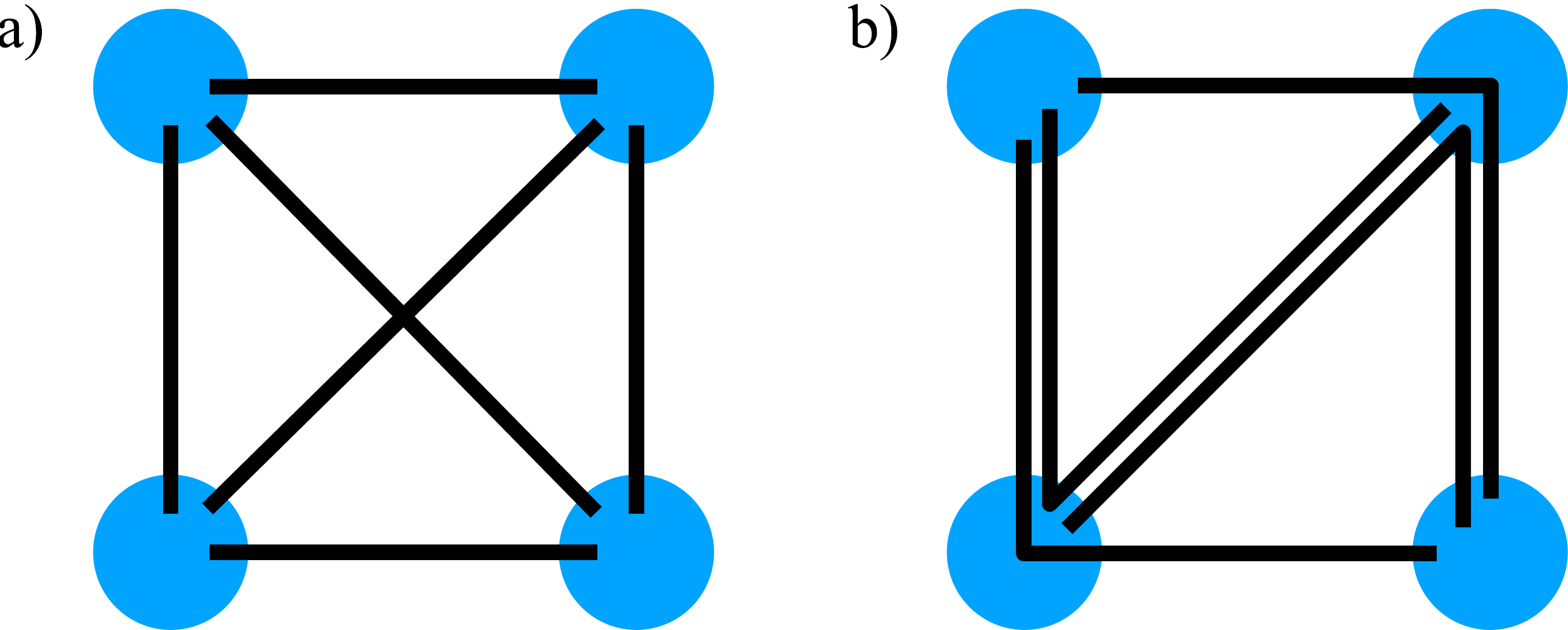}
\end{center}
\end{figure}

CGTNS is an approximation to a CAS configuration interaction (CAS-CI) wave function, {\it i.e.}, 
to FCI in a restricted orbital space. Higher accuracy can be achieved 
by introducing higher-order correlators\cite{chan2009}. For example, one may choose a tensor network of 3-site correlators,
\begin{equation}
\left| \Psi^{3s}_{\rm CGTNS} \right\rangle =\sum_{n_1 n_2 \ldots n_M} \prod_{i \le j \le k} C_{n_i n_j n_k}^{[ijk]} \left| n_1 n_2 \ldots n_M \right\rangle,
\end{equation}
where $\mathbf{C}^{[ijk]}$ is the third-order tensor
\begin{equation}\label{eq:3-site}
\mathbf{C}^{[ijk]} \equiv \left[
\begin{tikzpicture}[baseline={([yshift=-.5ex]current bounding box.center)},vertex/.style={anchor=base,
    circle,fill=black!25,minimum size=18pt,inner sep=2pt}]
  \matrix (m) [matrix of math nodes, row sep=0.25em,
    column sep=0.25em]{
    & C_{001}^{[ijk]}& & C_{011}^{[ijk]} \\
    C_{000}^{[ijk]} & & C_{010}^{[ijk]} & \\
    & C_{101}^{[ijk]} & & C_{111}^{[ijk]} \\
    C_{100}^{[ijk]} & & C_{110}^{[ijk]} & \\};
  \path[]
    (m-1-2) edge (m-1-4) edge (m-2-1)
            edge [densely dotted] (m-3-2)
    (m-1-4) edge (m-3-4) edge (m-2-3)
    (m-2-1) edge [-,line width=6pt,draw=white] (m-2-3)
            edge (m-2-3) edge (m-4-1)
    (m-3-2) edge [densely dotted] (m-3-4)
            edge [densely dotted] (m-4-1)
    (m-4-1) edge (m-4-3)
    (m-3-4) edge (m-4-3)
    (m-2-3) edge [-,line width=6pt,draw=white] (m-4-3)
            edge (m-4-3);
\end{tikzpicture}\right].
\end{equation}
In the limit of $M$-site correlators the coefficients of Eq.\ (\ref{eq:gen_state}), $C_{n_1 n_2 \ldots n_N}$, are recovered. In 
a 3-site correlator ansatz the total number of correlators (third-order tensors) is $1/6M(M+1)(M+2)$. As the tensor of third order, 
$\mathbf{C}^{[ijk]}$ of Eq.\ (\ref{eq:3-site}), has $q^3$ elements, this yields $1/6M(M+1)(M+2)q^3$ variational degrees of freedom. 
In analogy to the case of second-order tensors, one can remove {\it si} tensors of the type $\mathbf{C}^{[iii]}$, 
$\mathbf{C}^{[iik]}$, and $\mathbf{C}^{[ikk]}$ to obtain
\begin{equation}
\left| \Psi^{3s/si}_{\rm CGTNS} \right\rangle =\sum_{n_1 n_2 \ldots n_M} \prod_{i < j < k} C_{n_i n_j n_k}^{[ijk]} \mid n_1 n_2 \ldots n_M \rangle,
\end{equation}
which removes $M^2$ correlators. A graphical representation of such a tensor network is shown in Figure \ref{fig:graphs} b). As 
will be shown in Sections \ref{sec:sextet} and \ref{sec:doublet}, {\it si} correlators do {\it not} play a negligible role in the 3-site 
correlator ansatz in contrast to their 2-site correlator analogs. 

\subsubsection{Hybrid 2-Site and 3-Site Correlator Ans\"atze}
\begin{figure}[H]
\caption{Scaling of variational parameters in CAS-CI and various CGTNS parameterizations 
with increasing CAS($N$,$M_{\rm orb}$) sizes for a number of electrons $N$ identical to the number of active
spatial orbitals $M_{\rm orb}$, $N=M_{\rm orb}$. The colored shaded regions denote active spaces for which the low-order CGTNS ansatz of the 
same color introduces less variational parameters than the exact solution. \label{fig:scaling}}
\begin{center}
\includegraphics[scale=0.6]{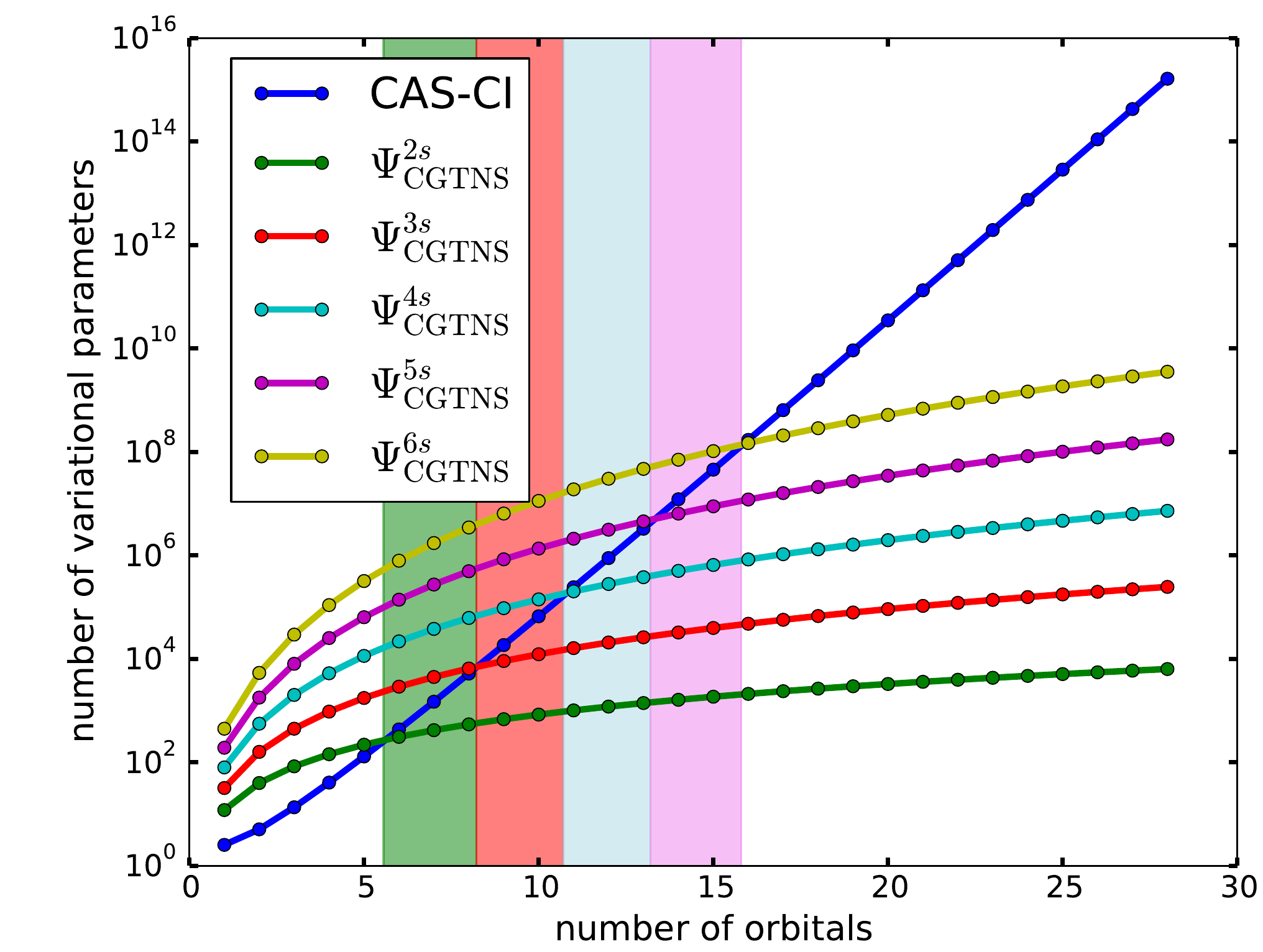}
\end{center}
\end{figure}

For better optimization efficiency, correlators can be introduced and optimized gradually starting with 2-site correlators 
and continuing with 3-site correlators. Having first optimized 2-site correlators, we may freeze their values and start the optimization of 
3-site correlators incorporated in the ansatz as scaling factors
\begin{equation}
\left| \Psi^{3s[2s]}_{\rm CGTNS} \right\rangle =\sum_{n_1 n_2 \ldots n_M} \underbrace{\prod_{i \le j} C_{n_i n_j}^{[ij]}}_{\rm frozen} 
\underbrace{\prod_{k \le l \le m} C_{n_k n_l n_m}^{[klm]}}_{\rm active} \left| n_1 n_2 \ldots n_M \right\rangle,
\end{equation}
or in self-interaction free form
\begin{equation}
\left| \Psi^{3s/si[2s]}_{\rm CGTNS} \right\rangle =\sum_{n_1 n_2 \ldots n_M} \underbrace{\prod_{i \le j} C_{n_i n_j}^{[ij]}}_{\rm frozen} 
\underbrace{\prod_{k < l < m} C_{n_k n_l n_m}^{[klm]}}_{\rm active} \left| n_1 n_2 \ldots n_M \right\rangle.
\end{equation}
However, introduction of products of 2-site and 3-site correlators will augment the non-linear structure of the CGTNS ansatz, which 
in turn could hamper the convergence of the optimization and increase the probability of getting trapped in local minima. This problem
could be alleviated by splitting up the hybrid ansatz into a {\it sum} of 2-site correlator and 3-site correlator products
\begin{equation}
\left| \Psi^{3s+[2s]}_{\rm CGTNS} \right\rangle =\sum_{n_1 n_2 \ldots n_M} \bigg[ \underbrace{\prod_{i \le j} C_{n_i n_j}^{[ij]}}_{\rm frozen} 
+ \underbrace{\prod_{k \le l \le m} C_{n_k n_l n_m}^{[klm]}}_{\rm active} \bigg] \left| n_1 n_2 \ldots n_M \right\rangle,
\end{equation}
or in self-interaction free form
\begin{equation}
\left| \Psi^{3s/si+[2s]}_{\rm CGTNS} \right\rangle =\sum_{n_1 n_2 \ldots n_M} \bigg[\underbrace{\prod_{i \le j} C_{n_i n_j}^{[ij]}}_{\rm frozen} 
+ \underbrace{\prod_{k < l < m} C_{n_k n_l n_m}^{[klm]}}_{\rm active}\bigg] \left| n_1 n_2 \ldots n_M \right\rangle.
\end{equation}
These parameterization strategies can be naturally continued to higher-order correlators. It is possible, but of no practical value, 
to finally include up to $M$-site correlators. Already four-site correlators will make the ansatz intractable, because of their 
sheer number. Hence, only the selective inclusion of higher-order correlators will be feasible. The number of variational 
parameters for all CGTNS variants depends only on the number of spin orbitals, $M$. By analogy to $\Psi^{2s}_{\rm CGTNS}$ and 
$\Psi^{3s}_{\rm CGTNS}$, one can define the ans\"atze for 4-, 5-, 6-site {\it etc.}~correlators to be denoted as $\Psi^{4s}_{\rm CGTNS}$, 
$\Psi^{5s}_{\rm CGTNS}$, $\Psi^{6s}_{\rm CGTNS}$ and so forth. Assuming that the number of electrons $N$ is growing 
with the number of spatial orbitals, $M_{\rm orb}=M/2$, and that they are equal, $N=M_{\rm orb}$, we obtain for the number of ONVs, $N_{\rm ONV}$, with spin 
projection $M_s=0$ for the active space of $N$ electrons in $M_{\rm orb}$ orbitals denoted as CAS($N$,$M_{\rm orb}$)\cite{molcas8}
\begin{equation}
N_{\rm ONV}=\frac{2}{\pi M_{\rm orb}}4^{M_{\rm orb}}.
\end{equation}
Note that if one exploits symmetry, this number can be decreased. The scaling of the number of variational parameters with 
respect to the number of orbitals for all variations of CGTNS as well as for CAS-based methods is shown in Figure \ref{fig:scaling}.

Clearly, one would stop the systematic extension of the CGTNS ansatz by higher-order correlators when the change in energy drops 
below a certain threshold. It would be most desirable to determine this threshold so that relative energies 
rather than absolute electronic energies are accurately approximated.
The maximum number of variational parameters would then be determined by the 
highest-order tensor network ansatz. An alternative strategy is to introduce higher-order correlators at or before the points where the 
curve corresponding to a specific CGTNS ansatz crosses the CAS-CI curve in Figure~\ref{fig:scaling}. The colored regions in 
Figure~\ref{fig:scaling} show the scope of application for each CGTNS parameterization (the regions are marked with a color corresponding 
to the curve representing that CGTNS ansatz). Such a strategy, however, only gives a qualitative idea on the applicability 
of CGTNS schemes. A more rigorous way will be developed in Section \ref{sec:measure}.

The set of higher-order correlators can inflate the variational space to that of CAS-CI and beyond (the regions where curves corresponding to CGTNS 
ans\"atze are above the CAS line in Figure \ref{fig:scaling}). To properly cope with such situations requires to introduce 
higher-order correlators only for certain sites which may be determined based on entanglement measures. The entanglement between 
sites can be estimated from single-orbital entropies and mutual information entropies\cite{lege2003,lege2004,riss2006,bogu2012b,stei2016} 
obtained for the low-order correlators such as those in $\Psi^{2s}_{\rm CGTNS}$. In this, work such an ansatz 
will be denoted as $\Psi_{\rm CGTNS}^{3s[2s]{\rm sel}}$. 

\subsection{The Spin-Adapted CGTNS Ansatz}
The expectation value of the Hamiltonian operator over an approximate $N$-electron wave function $\Psi_{\rm CGTNS}$ 
is an upper bound to the exact CAS-CI reference energy,
\begin{equation}
E_{\rm CAS\text{-}CI}=\frac{\left\langle \Psi_{\rm CAS\text{-}CI} \right| H \left| \Psi_{\rm CAS\text{-}CI} \right\rangle}{\left\langle \Psi_{\rm CAS\text{-}CI} 
| \Psi_{\rm CAS\text{-}CI} \right\rangle} \le \frac{\left\langle \Psi_{\rm CGTNS} \right| H \left| \Psi_{\rm CGTNS} \right\rangle}
{\left\langle \Psi_{\rm CGTNS} | \Psi_{\rm CGTNS} \right\rangle},
\end{equation}
where for $\Psi_{\rm CGTNS}$ we can have any approximation introduced above ($\Psi_{\rm CGTNS}^{2s}$, $\Psi_{\rm CGTNS}^{3s}$, 
$\Psi_{\rm CGTNS}^{2s/si}$, $\Psi_{\rm CGTNS}^{3s/si}$, $\Psi_{\rm CGTNS}^{3s/si[2s]}$, $\Psi_{\rm CGTNS}^{3s[2s]{\rm sel}}$). 
For a specific ONV $\left| t \right\rangle = \left| t_1 t_2 \ldots t_M \right\rangle$ in the $\Psi_{\rm CGTNS}^{2s}$ 
ansatz, for instance, we approximate a CI coefficient $C_{t}=C_{t_1 t_2 \ldots t_M}$ as,
\begin{equation}\label{eq:weights}
C_{t} \approx C_{t}^{\rm CGTNS} = \langle t \left| \Psi_{\rm CGTNS} \right\rangle = \prod_{i \le j}C_{n_i n_j}^{[ij]}, \quad \forall n_i, n_j \in \left|t\right\rangle,
\end{equation}
so that the CGTNS wave function can be rewritten in compact form as
\begin{equation}\label{eq:ci_like}
\left| \Psi_{\rm CGTNS} \right\rangle = \sum_{n} C_{n}^{\rm CGTNS} \left| n \right\rangle.
\end{equation}
Then, the normalization condition reads
\begin{align}\label{eq:norm_cgtns}
\left\langle \Psi_{\rm CGTNS} | \Psi_{\rm CGTNS} \right\rangle &=\sum_{nl}\left(C^{\rm CGTNS}_{n}\right)^{*}C^{\rm CGTNS}_{l}\langle n \left| l \right\rangle\\ \nonumber
&=\sum_{nl}\left(C_{n}^{\rm CGTNS}\right)^{*}C^{\rm CGTNS}_{l}\delta_{nl}
=\sum_{n}\left(C_{n}^{\rm CGTNS}\right)^{2}.
\end{align}
Accurate calculations demand spin-adapted configuration state functions (CSFs), 
\begin{equation}\label{eq:csf}
\left| \Phi^{\rm CSF}_{p} \right\rangle = \sum_{n} K_{pn} \left| n \right\rangle,
\end{equation}
where $K_{pn}$ are Clebsch--Gordan coefficients generating these spin-adapted basis functions. The CAS-CI function 
can then be expanded as
\begin{equation}\label{eq:sa}
\left| \Psi_{\rm CAS\text{-}CI} \right\rangle = \sum_{p} S_{p} \left| \Phi^{\rm CSF}_{p} \right\rangle.
\end{equation}
Since Clebsch--Gordan coefficients $K_{pn}$ from Eq.~(\ref{eq:csf}) are fixed, one can no longer optimize the weights of 
Slater determinants, $C_{n}^{\rm CGTNS}$, and the straightforward CGTNS concept breaks down. However, it is possible to 
approximate $S_{p}$ in the previous equation as a sum of Clebsch--Gordan coefficients $K_{pn}$ scaled by $C_{n}^{\rm CGTNS}$ 
for every $\mid n \rangle$ in Eq.\ (\ref{eq:csf}),
\begin{equation}\label{eq:st}
S_{p} \approx  S_{p}^{\rm CGTNS}= \sum_{n} K_{pn} C_{n}^{\rm CGTNS}.
\end{equation}
With the weights $S_{p}^{\rm CGTNS}$ in Eq.\ (\ref{eq:st}), we define the spin-adapted CGTNS ansatz as
\begin{equation}\label{eq:csf_like}
\mid \Psi_{\rm CGTNS} \rangle = \sum_{p} \sum_{n} K_{pn} C_{n}^{\rm CGTNS} \mid \Phi^{\rm CSF}_{p} \rangle
= \sum_{p} S_{p}^{\rm CGTNS} \mid \Phi^{\rm CSF}_{p} \rangle .
\end{equation}
The normalization will then take the following form
\begin{align}\label{eq:csf_norm}
\langle \Psi_{\rm CGTNS} \mid \Psi_{\rm CGTNS} \rangle &= \sum_{pn} \sum_{ql} S_p^{\rm CGTNS} K_{pn} S_q^{\rm CGTNS} K_{ql} \langle n \mid l \rangle\\ \nonumber
&= \sum_{pn} \sum_{ql} S_p^{\rm CGTNS} K_{pn}  S_q^{\rm CGTNS} K_{ql} \delta_{nl} \\ \nonumber
&= \sum_{pq} S_p^{\rm CGTNS} S_q^{\rm CGTNS} \sum_{n} K_{pn} K_{qn},
\end{align}
where we assume real coefficients $S_{p}^{\rm CGTNS}$ and $K_{pn}$. In the following, we will always consider 
spin-adapted CGTNS parameterizations.

\subsection{Monte Carlo Optimization}
The highly nonlinear dependence of the CGTNS ansatz on correlators 
makes convergence of optimization procedures toward a global minimum a difficult task. Hence, we continue to employ a variational 
Monte Carlo optimization scheme\cite{mart2010b,sand2007}. With Eqs.\ (\ref{eq:csf_like}) and (\ref{eq:csf_norm}), the 
expectation value of the Hamiltonian for a CGTNS wave function reads
\begin{equation}\label{eq:approx}
E_{\rm CGTNS}=\frac{\sum_{r} S_{r}^{\rm CGTNS} \langle \Psi_{\rm CGTNS} \mid H \mid \Phi_{r}^{\rm CSF} \rangle }{\sum_{pq} S_p^{\rm CGTNS} S_q^{\rm CGTNS} \sum_{n} K_{pn} K_{qn}}.
\end{equation}
We can rewrite Eq.\ (\ref{eq:approx}) in a more useful form for Monte Carlo sampling,
\begin{equation}
E_{\rm CGTNS}=\frac{\sum_{r}\left(S_{r}^{\rm CGTNS}\right)^{2} \displaystyle \frac{\langle \Psi_{\rm CGTNS} \mid H \mid \Phi_{r}^{\rm CSF} \rangle}{S_{r}^{\rm CGTNS}} }
{\sum_{pq} S_p^{\rm CGTNS} S_q^{\rm CGTNS} \sum_{n} K_{pn} K_{qn}},
\end{equation}
where the $\left(S_{r}^{\rm CGTNS}\right)^{2}$ represent strictly non-negative probabilities for corresponding energy estimators
\begin{equation}\label{eq:est}
E_r=\frac{\left\langle \Psi_{\rm CGTNS} \left| H \right| \Phi_{r}^{\rm CSF} \right\rangle}{S_{r}^{\rm CGTNS}}= \sum_{s} \frac{S_{s}^{\rm CGTNS}}{S_{r}^{\rm CGTNS}}\left\langle 
\Phi_{s}^{\rm CSF} \left| H \right| \Phi_{r}^{\rm CSF} \right\rangle.
\end{equation}
Since every $S_{s}^{\rm CGTNS}$ is determined by a set of correlators $\mathbf{\tilde{C}}$, with 
$\mathbf{\tilde{C}}$ = $\{\mathbf{C}^{[11]}, \mathbf{C}^{[12]},$ $ \dots,$ $ \mathbf{C}^{[ij]},$ $ \dots,$ $ \mathbf{C}^{[NN]}\}$ through Eqs.~(\ref{eq:weights}) 
and (\ref{eq:st}) for the 2-site case, for every choice of correlators $\mathbf{\tilde{C}}$ one can assign an energy $E_r(\mathbf{\tilde{C}})$ 
\begin{equation}\label{eq:ener_estimate}
E_r(\mathbf{\tilde{C}})= \sum_{s} \frac{S_{s}^{\rm CGTNS}(\mathbf{\tilde{C}})}{S_{r}^{\rm CGTNS}(\mathbf{\tilde{C}})}\left\langle \Phi_{s}^{\rm CSF} 
\left| H \right| \Phi_{r}^{\rm CSF} \right\rangle.
\end{equation}
Introducing an artificial temperature $T$ (a parameter with the dimension of energy, measured in Hartree), it is possible to 
sample the continuous variables $\mathbf{\tilde{C}}$ following a canonical ensemble with the weight of a configuration given 
by $\exp{\left[-E_r(\mathbf{\tilde{C}})/T\right]}$. The limit $T\rightarrow 0$ Hartree yields the desired ground state of the 
molecule. The optimization procedure can be easily controlled by tuning $T$. To avoid getting trapped in local minima, the parallel 
tempering scheme is applied\cite{mart2010b} during the optimization with swap-move probabilities between two neighboring 
temperatures defined as
\begin{equation}\label{eq:swap_prob}
p((T_i,E_i)\leftrightarrow(T_{i+1},E_{i+1}))={\rm min}\{1,\exp{(\Delta E/ \Delta T)}\},
\end{equation}
where $\Delta E = E_{i+1} - E_i$ and $\Delta T = T_{i+1}T_i/(T_i-T_{i+1})$. The set of $P$ temperatures 
in the range $[T_1,T_P]$ are chosen according to the formula
\begin{equation}
T_{l}=T_{1}\left( \exp{\frac{\ln{T_{P}} - \ln{T_{1}}}{P-1}} \right)^{l-1}, \mbox{ with } l=1 \ldots P.
\end{equation}
It is clear from Eq.~(\ref{eq:ener_estimate}) that this procedure will only be feasible for large CAS if not all $\Phi_{r}^{\rm CSF}$ 
are required. The exponential scaling of the dimension of the Hilbert space with the number of orbitals is clearly a restriction 
of the CGTNS approach which it shares with the FCI Quantum Monte Carlo approach of Alavi and co-workers\cite{boot2009,shep2012,boot2013}.
Accordingly, CSFs that hardly contribute to the energy must be omitted.

\subsubsection{Gradient-based Optimization}\label{sec:grad_opt}
The Monte Carlo optimized CGTNS ansatz, {\it e.g.} $\Psi^{2s}_{\rm CGTNS}$, can be taken as a starting point for a non-stochastic 
local optimization for refinement. The local gradient
\begin{align}\label{eq:grad}
\nabla E_{\rm CGTNS}&= \left( \frac{\partial E_{\rm CGTNS}}{\partial C^{[11]}_{00}}, \frac{\partial E_{\rm CGTNS}}{\partial C^{[11]}_{01}}, 
\frac{\partial E_{\rm CGTNS}}{\partial C^{[11]}_{10}}, \frac{\partial E_{\rm CGTNS}}{\partial C^{[11]}_{11}}, \ldots,
\frac{\partial E_{\rm CGTNS}}{\partial C^{[MM]}_{11}}\right),
\end{align}
can be evaluated and exploited in such a local search. Introducing the correlators together with the corresponding gradient, 
Eq.~(\ref{eq:grad}), into the Quasi-Newton optimization method with the Broyden-Fletcher-Goldfarb-Shanno algorithm for an 
update of the Hessian matrix\cite{num_rep2007}, we can further lower the CGTNS energy. An alternative way is to consider 
only the gradient of the CGTNS energy for the correlators corresponding to certain 
pairs of sites $i$ and $j$ at a time
\begin{align}\label{eq:grad_site}
\nabla^{[ij]} E_{\rm CGTNS} &= \left( \frac{\partial E_{\rm CGTNS}}{\partial C^{[ij]}_{00}}, \frac{\partial E_{\rm CGTNS}}{\partial C^{[ij]}_{01}}, 
\frac{\partial E_{\rm CGTNS}}{\partial C^{[ij]}_{10}}, \frac{\partial E_{\rm CGTNS}}{\partial C^{[ij]}_{11}}\right).
\end{align}
Switching between all possible pairs of the sites in the CGTNS ansatz, convergence should be reached at some point. 
We will refer to the first optimization strategy as a gradient optimization, while the second strategy will 
be denoted ``reduced'' gradient optimization in the following. Following Chan and coworkers\cite{chan2009}, we 
note that, because the wave function is linear with respect to correlators of a given set of sites, the components 
of the gradient for the chosen correlator may define a vector space for the optimization. Hence, it is possible 
to introduce the Hamiltonian and overlap matrix in the vector space spanned by the components of the local 
gradient\cite{chan2009}. Finding the eigenpairs corresponding to this Hamiltonian and switching between all possible 
pairs, we may finally obtain the same solution as in the case of ``reduced'' gradient optimization. 

\section{Computational Details}\label{sec:comp_det}
A suitable molecule for the analysis of the various CGTNS parameterizations should have a strong multi-configurational 
character. Manganocene is such a molecule. It is particularly interesting because most DFT calculations fail to predict 
the proper ground spin state\cite{salo2002}. While it is decisive to be able to predict the energy difference between 
high- and low-spin states in transition metal compounds, it turned out to be hard to predict the energy difference 
between sextet and doublet state in manganocene\cite{salo2002}. Satisfactory accuracy can be achieved by describing 
these two spin states with the CASPT2 method\cite{phun2012}. Phung {\it et al.}~showed\cite{phun2012} that dynamic 
electron correlation is as important as static electron correlation for the quantitative prediction of the spin state 
splitting (see also Ref.\ \citenum{stei16b} for an in-depth discussion). 
Our CGTNS ansatz operates in an active space of selected orbitals and hence approximates a CAS-CI wave 
function. While this is no issue for the analysis of the CGTNS parameterizations, reliable predictions will require 
to consider dynamic correlation, which can, for instance, be included by short-range DFT\cite{from2007,hede2015}. 
Hence, the theoretical reference for our study will be the CAS-SCF result rather than the CASPT2 data by Phung 
{\it et al.}\cite{phun2012}.

Density Functional Theory (DFT) structure optimizations of manganocene were conducted with the TURBOMOLE program version 
6.5\cite{TURBOMOLE6.5,ahlr1989} in the doublet and sextet states. The hybrid Perdew--Burke--Ernzerhof (PBE0)\cite{perd1996b} 
density functional together with triple-$\zeta$ valence polarized (def2-TZVP)\cite{weig1998} (for carbon and hydrogen atoms) 
and quadruple-$\zeta$ valence polarized (def2-QZVPP)\cite{weig2003} (for the manganese atom) basis sets were chosen. In addition, 
single-point DFT calculations were performed for the sextet and doublet states with the pure Perdew--Burke--Ernzerhof 
(PBE)\cite{perd1996} density functional. Note, that for all DFT calculations Grimme D3 dispersion corrections\cite{grim2010} 
and the second-order scalar-relativistic Douglas-Kroll-Hess Hamiltonian\cite{hess1986,wolf2002,reih2004c,peng2013} were switched on. 

\begin{figure}[H]
\caption{The PBE0-D3/def2-TZVP(C,H)/def2-QZVPP(Mn) structures of man\-ga\-no\-cene in the doublet state (left) 
and the sextet state (right). Hydrogen atoms in white, carbon atoms in black, and manganese atoms in purple.
\label{fig:structure}}
\begin{center}
\includegraphics[scale=0.6]{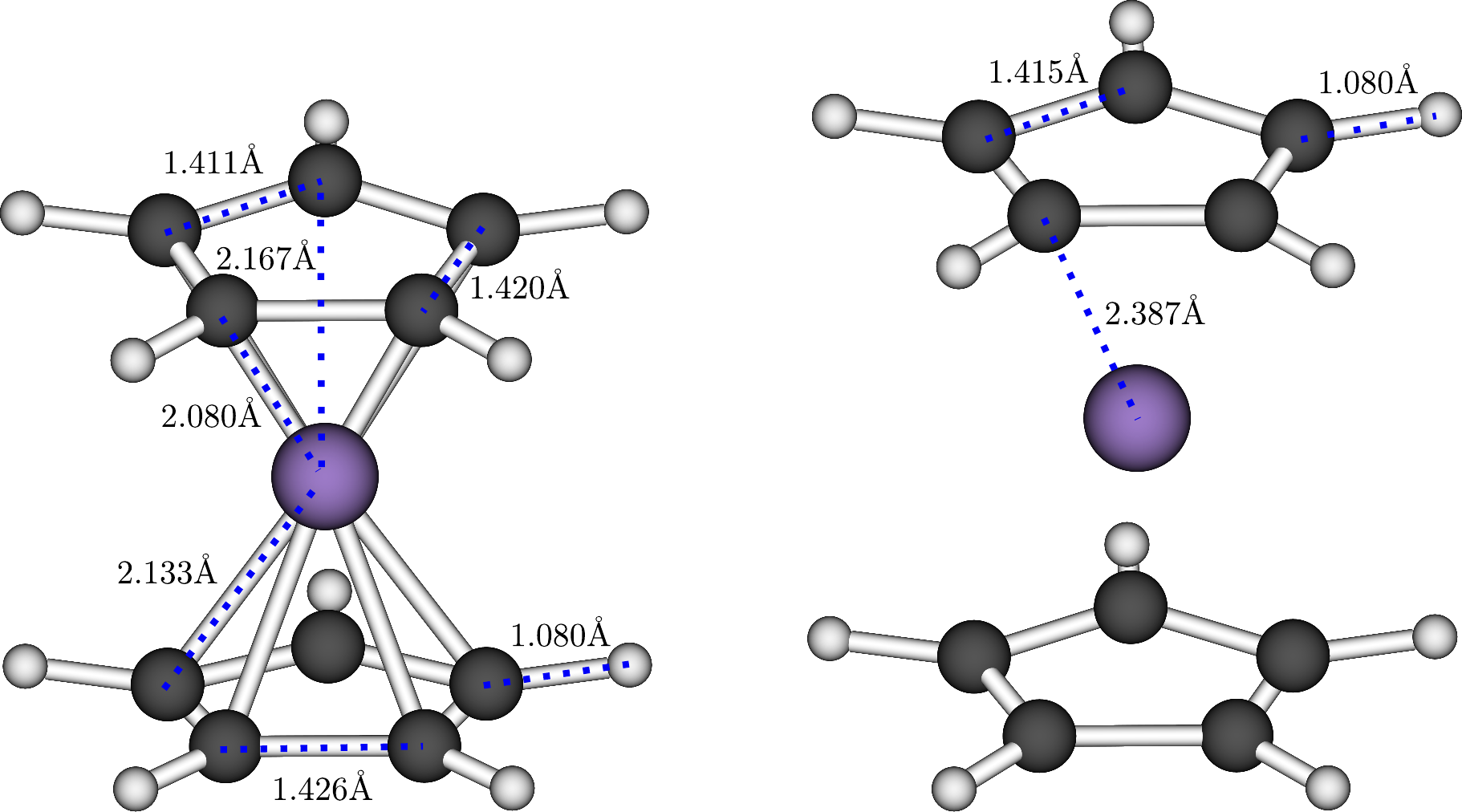}	
\end{center}
\end{figure}

The doublet and sextet manganocene structures were optimized with DFT-PBE0-D3 in $C_{2v}$ and $D_{5h}$ symmetries, respectively. 
These optimized structures were then taken for all single-point calculations in this work; see Figure \ref{fig:structure}.

All reference CAS-SCF calculations were performed with the MOLCAS 8.1 package\cite{molcas8}. The extended 
ANO-RCC basis sets (with a total of 1487 basis functions) were employed with [6s4p3d1f] contraction for 
hydrogen\cite{widm1990}, [8s8p4d3f2g] contraction for carbon\cite{roos2004}, and [10s9p8d6f4g2h] contraction
for manganese\cite{roos2005}. Two-electron integrals were approximated with a Cholesky decomposition 
technique\cite{aqui2007} using a threshold of 10$^{-6}$ Hartree. Also in these calculations, 
the second-order scalar-relativistic Douglas--Kroll--Hess Hamiltonian\cite{hess1986,wolf2002,reih2004c,peng2012a} 
was chosen. 

The CGTNS program reads one- and two-electron molecular orbital integrals for the second-quantized electronic Hamiltonian generated 
by MOLCAS. The integrals are calculated from the natural orbitals of the corresponding CAS-SCF reference calculations. For the CGTNS 
calculations we improved on our original implementation presented in Ref.\ \citenum{mart2010b}. The gradient optimization and the ``reduced'' 
gradient optimization according to Eq.\ (\ref{eq:grad}) and Eq.\ (\ref{eq:grad_site}), respectively (see Section~\ref{sec:grad_opt}), were 
implemented. In addition to the existing $\Psi^{2s}_{\rm CGTNS}$ ansatz, we implemented $\Psi^{2s/si}_{\rm CGTNS}$, $\Psi^{3s}_{\rm CGTNS}$, 
$\Psi^{3s/si}_{\rm CGTNS}$, $\Psi^{3s[2s]}_{\rm CGTNS}$, $\Psi^{3s+[2s]}_{\rm CGTNS}$, $\Psi^{3s/si[2s]}_{\rm CGTNS}$, and 
$\Psi^{3s/si+[2s]}_{\rm CGTNS}$. 

The $\Psi_{\rm CGTNS}^{3s[2s]{\rm sel}}$ ansatz is a slightly modified version of the $\Psi^{3s[2s]}_{\rm CGTNS}$ 
parameterization. In contrast to $\Psi^{3s[2s]}_{\rm CGTNS}$, it contains the 3-site correlators corresponding only to the most entangled spin 
orbitals (the same holds true for $\Psi^{3s+[2s]{\rm sel}}_{\rm CGTNS}$ and $\Psi^{3s+[2s]}_{\rm CGTNS}$). Ideally, these selected spin orbitals 
should be chosen using entanglement measure based on the single-orbital entropies and the mutual information\cite{lege2003,lege2004,riss2006,bogu2012b,stei2016}. 
However, here we select 3-site correlators for spin orbitals corresponding to spatial orbitals with natural orbital occupation numbers 
in the range $[0.02,1.98]$ (unless otherwise noted). This follows the Unrestricted Natural Orbital-CAS (UNO-CAS) 
model\cite{pula1988,bofi1989}, in which the active space is constructed from Unrestricted Hartree--Fock (UHF) natural (spatial) orbitals with occupation 
numbers between 0.02 and 1.98. 
The same strategy was also adopted for choosing an active space in DMRG calculations\cite{kell2015}. 
UHF natural orbitals may even represent a good choice for the orbital basis\cite{pula1988,bofi1989,kell2015}.

\section{Discussion}
\subsection{Reference CAS-SCF(9,12) Energy Difference}
\begin{figure}[H]
\caption{Converged natural orbitals 
--- denoted as \textbf{D-i} and \textbf{S-i} (in red color) for the doublet and sextet states, respectively ($\mathbf{i=\{1,2,3,\dots,12\}}$) ---
that constitute the active spaces in the CAS-SCF(9,12) reference calculations of manganocene.
\label{fig:orbitals}}
\begin{center}
\includegraphics[scale=0.525]{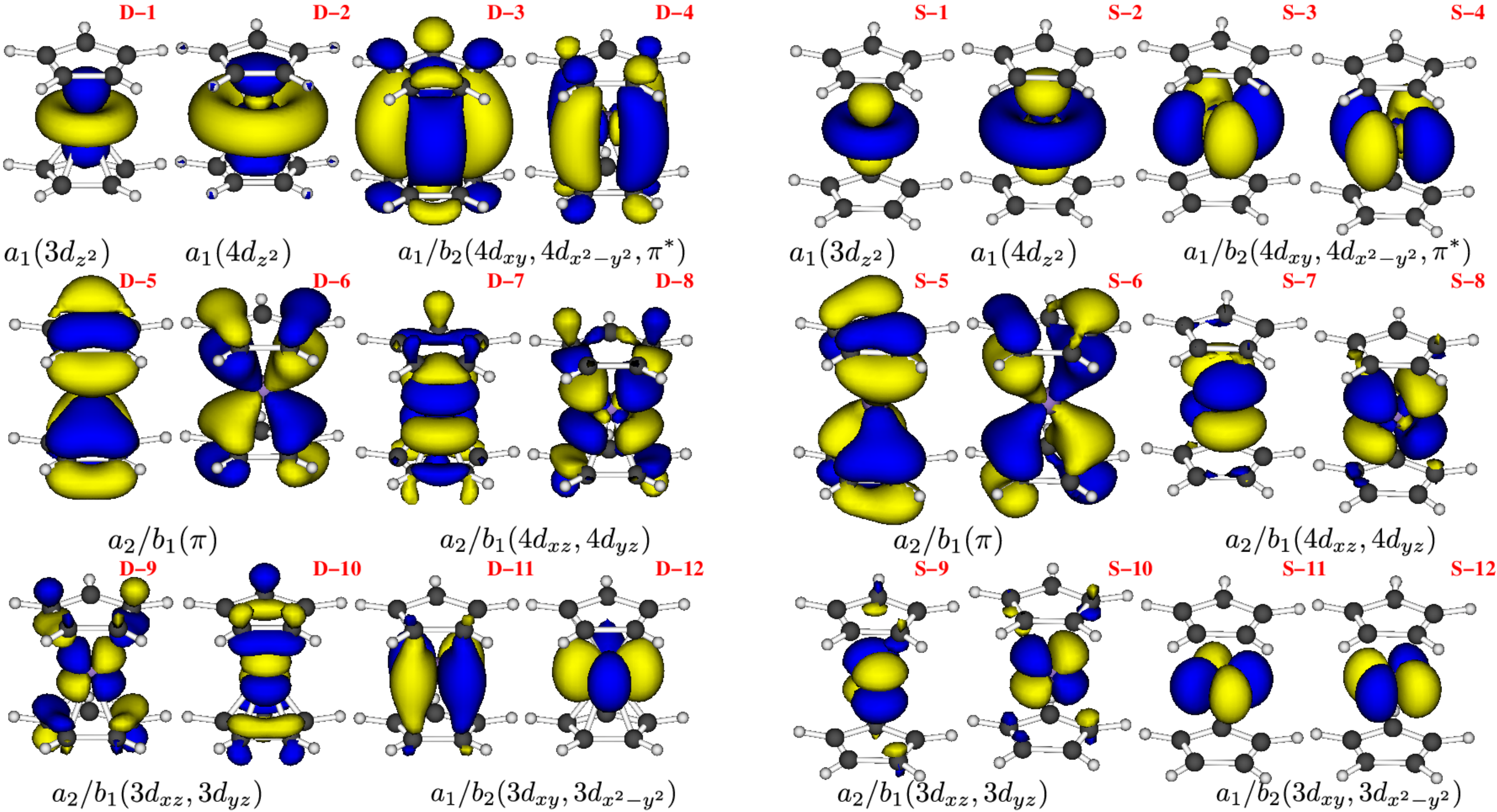}	
\end{center}
\end{figure}

The CAS-SCF(9,12) reference calculations for both spin states in $C_{2v}$ symmetry were carried out for active orbital spaces proposed
in Ref.\ \citenum{phun2012}. The orbitals included into the active spaces of the CAS-SCF calculation are shown in Figure \ref{fig:orbitals}.

Note that an active space consisting of 9 electrons in 12 spatial orbitals chosen according to Ref.\ \citenum{phun2012} 
turned out to be not without difficulties. In particular, for both spin states one cannot obtain pure $\pi^*$ orbitals without 
the inclusion of lower lying $\pi$ orbitals, high lying $\pi^*$ orbitals, and $4d_{xy}$/$4d_{x^2-y^2}$ 
orbitals into the active space. This would introduce two additional electrons and six additional orbitals into active space. Moreover, symmetry breaking may 
occur for the sextet state. Nevertheless, in order to be in line with Ref.\ \citenum{phun2012}, we adopt the smaller active space from 
that reference as the choice of the active space is of little importance 
for our wave-function parameterization analysis. The natural orbitals produced in the CAS-SCF(9,12) calculation were then chosen for 
the CGTNS calculations. Hence, the CGTNS result approximates the CAS-SCF(9,12) result.

The reference energy difference calculated with CAS-SCF(9,12) for the sextet and doublet states is presented in Table \ref{tab:ref_energy}. 
The CAS-SCF(9,12) energy difference of -40.59 kcal/mol, deviates from the experimental value of 3.58 kcal/mol because of the lack of
dynamic electron correlation. The PBE0 result of
-8.28 kcal/mol (see DFT-PBE0-D3 in Table \ref{tab:ref_energy}) is much closer to the experimental result, but fails to predict 
the correct sign. PBE (see DFT-PBE-D3 in Table \ref{tab:ref_energy}), predicts the correct sign, but the energy is overestimated 
by approximately 19 kcal/mol. The CASPT2 energy difference (see CASPT2 in Table \ref{tab:ref_energy}) is very close to the 
experimental result\cite{phun2012}. We emphasize again that, in this work, we concentrate on the assessment of static correlation described 
by tensor network parameterizations. Dynamic correlation may be included through second order perturbation theory\cite{ande1992,ange2001} 
or short-range DFT methods\cite{from2007,hede2015}.

\begin{table}[H]
\caption{Electronic energy difference (in kcal/mol) of manganocene in the sextet $E[^6A_1]$ and doublet $E[^2A_1]$ states, $E[^6A_1]-E[^2A_1]$
obtained for different methods described in Section \ref{sec:comp_det}. 
CASPT2 and experimental (exp.) results were taken from Ref.~\citenum{phun2012}.}\label{tab:ref_energy}
\begin{center}
\begin{tabular}{lccccc}
\hline\hline
 & DFT-PBE-D3 & DFT-PBE0-D3     & CAS-SCF(9,12) & CASPT2 & exp.\ \\
\hline
$E[^6A_1]-E[^2A_1]$ &        22.55 &        $-8.28$ &       $-40.59$ & 5.77 & 3.58\\
\hline\hline
\end{tabular}
\end{center}
\end{table}

\subsection{Manganocene --- Sextet State}\label{sec:sextet}
\begin{figure}[H]
\caption{Convergence behavior of $\Psi_{\rm CGTNS}^{2s}$ and $\Psi_{\rm CGTNS}^{2s/si}$ parameterizations for manganocene 
in the lowest sextet state.
\label{fig:conv_sext}}
\begin{center}
\includegraphics[scale=0.6]{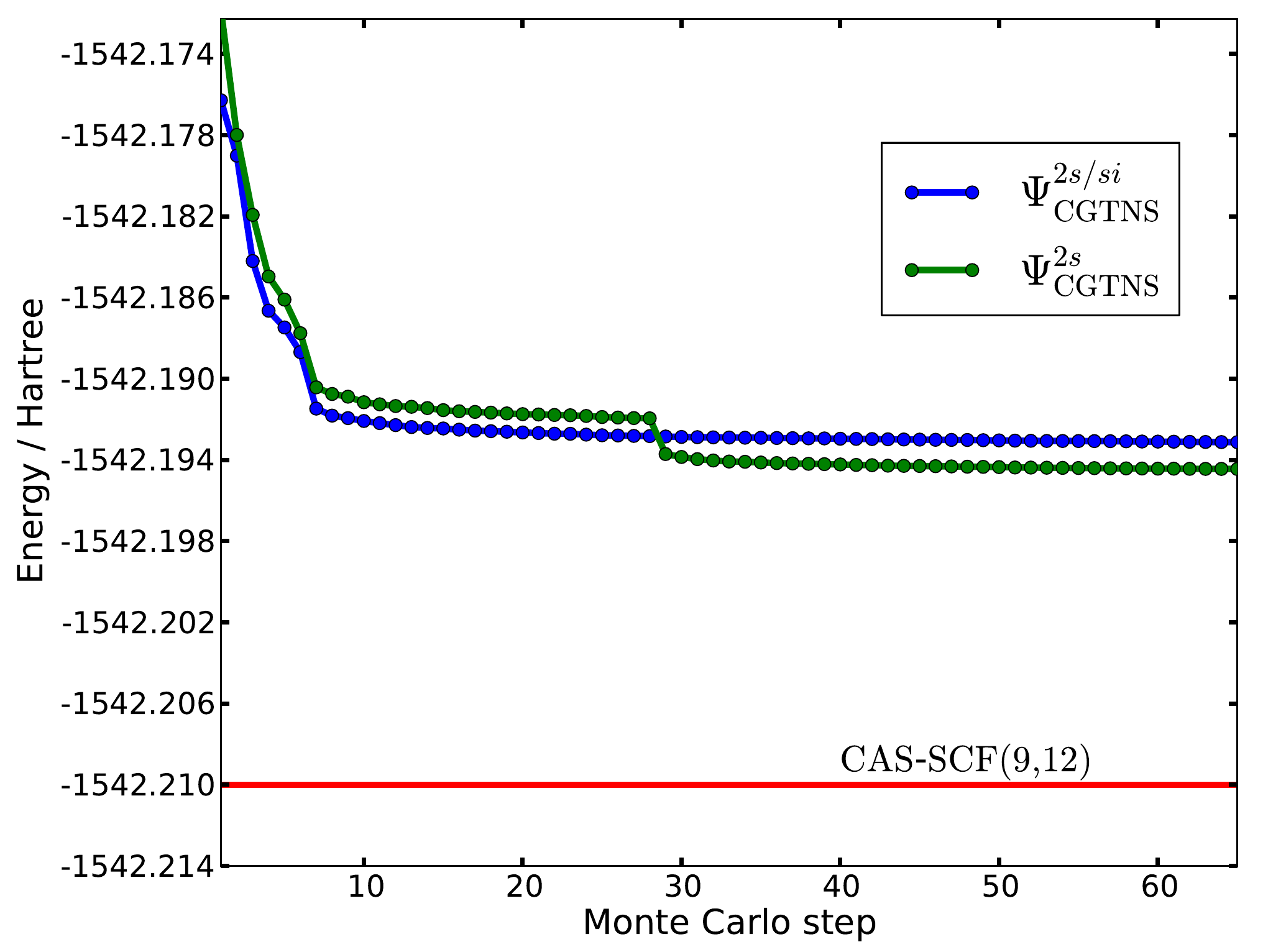}	
\end{center}
\end{figure}

13108 ONVs span the configurational space corresponding to an active space of 9 electrons in 12 spatial orbitals 
(see Figure~\ref{fig:orbitals}) for the sextet state of manganocene in $C_{2v}$ point group symmetry. The number 
of CGTNS variational parameters for 24 spin orbitals is 1200 for $\Psi_{\rm CGTNS}^{2s}$ and 1104 for $\Psi_{\rm CGTNS}^{2s/si}$. 
Hence, the parameterization in $\Psi_{\rm CGTNS}^{2s}$ and $\Psi_{\rm CGTNS}^{2s/si}$ reduce the variational space 
by more than 90\% in both cases, see Table \ref{tab:sextuplet}. The number of spin-adapted CSFs for the sextet state 
is 11628 and therefore not much lower than the total number of ONVs so that the CGTNS reduction is still approximately 
90\%. At the same time, the deviation of the CGTNS energy from the reference energy is lower than 20 mHartree, 
see Table \ref{tab:sextuplet}.
\begin{table}[H]
\caption{Electronic energies for the sextet state of manganocene. A positive/negative percentage indicates a 
decreased/increased parameter space compared to the 13108 CI coefficients in the CAS-SCF(9,12) reference calculation.}
\label{tab:sextuplet}
\begin{center}
\begin{tabular}{lrrr}
\hline\hline
parameterization & parameters & percentage & energy/Hartree\\
\hline
 CAS-SCF(9,12) 				& 13108 &         & $-1542.209620$ \\
 $\Psi_{\rm CGTNS}^{2s}$ 		& 1200	& 91\%   & $-1542.194072$ \\
 $\Psi_{\rm CGTNS}^{2s/si}$     	& 1104	& 92\%   & $-1542.192755$ \\
 $\Psi_{\rm CGTNS}^{3s/si}$             & 16192 &$-29$\% & $-1542.195739$ \\
 $\Psi_{\rm CGTNS}^{3s}$                & 20800 &$-59$\% & $-1542.197777$ \\
 $\Psi_{\rm CGTNS}^{3s/si[2s]}$         & 16192 &$-29$\% & $-1542.195415$ \\
 $\Psi_{\rm CGTNS}^{3s/si+[2s]}$        & 16192 &$-29$\% & $-1542.195290$ \\
 $\Psi_{\rm CGTNS}^{3s[2s]}$            & 20800 &$-59$\% & $-1542.195283$ \\
 $\Psi_{\rm CGTNS}^{3s+[2s]}$           & 20800 &$-59$\% & $-1542.195227$ \\
 $\Psi_{\rm CGTNS}^{3s[2s]{\rm sel}}$   & 4480  & 66\%   & $-1542.194826$ \\
 $\Psi_{\rm CGTNS}^{3s+[2s]{\rm sel}}$  & 4480  & 66\%   & $-1542.194822$ \\
\hline\hline
\end{tabular}
\end{center}
\end{table}
Note that the $\Psi_{\rm CGTNS}^{2s}$ parameterization is slightly more accurate than the $\Psi_{\rm CGTNS}^{2s/si}$ parameterization 
yielding a 0.001317 Hartree lower energy, induced by only 96 additional variational parameters. The convergence behavior of both 
parameterization is similar, see Figure \ref{fig:conv_sext}. The Monte Carlo optimizations for both CGTNS parameterizations have 
reached convergence. However, the Monte Carlo sampling of CGTNS parameters requires significantly more computational time than 
the traditional diagonalization approach. Hence, the CGTNS ansatz will be beneficial only for cases where exact diagonalization 
is no longer feasible.
As the 2-site correlator CGTNS energy deviates from the CAS-SCF reference, it is important to analyze whether 3-site correlators better 
approximate the reference result. The 3-site correlator variants of CGTNS $\Psi_{\rm CGTNS}^{3s}$/$\Psi_{\rm CGTNS}^{3s[2s]}$ and 
$\Psi_{\rm CGTNS}^{3s/si}$/$\Psi_{\rm CGTNS}^{3s/si[2s]}$ dramatically increase the variational space from 1200 to 20800 and 16192 
parameters, respectively (see Table \ref{tab:sextuplet}). Hence, it is inevitable to include only those 3-site correlators of the most 
entangled spin orbitals. According to the selection criterion described in Section~\ref{sec:comp_det}, we have chosen the natural 
orbitals \textbf{S-1}, \textbf{S-9}, \textbf{S-10}, \textbf{S-11}, \textbf{S-12}, see Figure~\ref{fig:orbitals}. The additional 
3-site correlators were constructed for 10 spin orbitals attributed to the selected natural orbitals, which resulted in a total 
of 1760 variational parameters for the hybrid CGTNS ansatz $\Psi_{\rm CGTNS}^{3s[2s]{\rm sel}}$. With this parameterization the 
energy converges very fast, but lowers it by only -0.272 mHartree. 

\begin{figure}[H]
\caption{Convergence behavior of 3-site correlator CGTNS parameterizations for manganocene in the lowest-energy sextet state.
\label{fig:conv_sext3}}
\begin{center}
\includegraphics[scale=0.6]{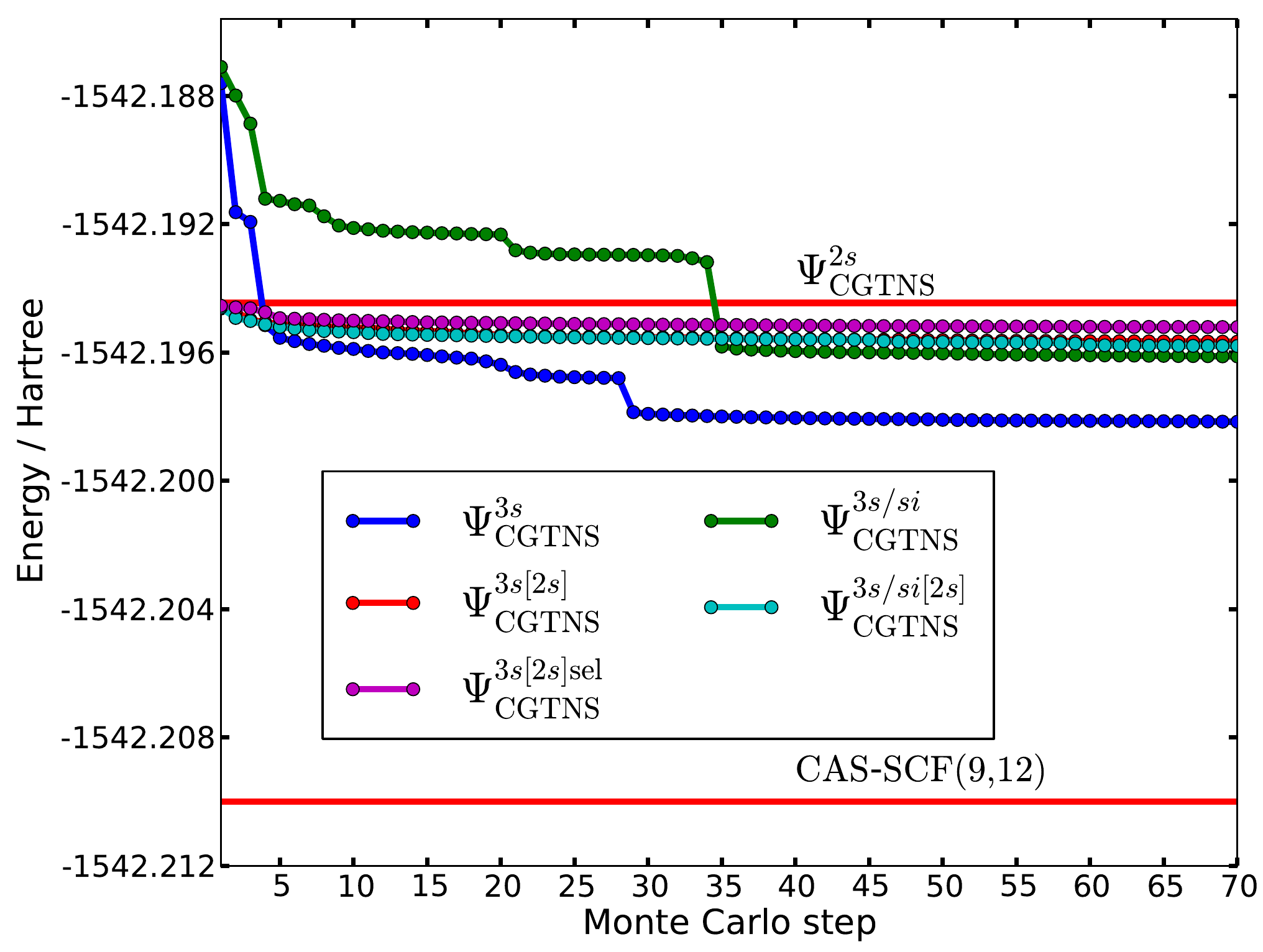}	
\end{center}
\end{figure}

If additional 3-site correlators are constructed considering also 
orbitals \textbf{S-5} and \textbf{S-6} (Figure~\ref{fig:orbitals}) that feature occupation numbers larger than 1.98 (i.e, 1.99 and larger), 
a total of 4480 
variational parameters results. In this case, the energy is lowered by 0.754 mHartree, see Table \ref{tab:sextuplet} and Figure \ref{fig:conv_sext3}. 
In order to understand the reason for this small effect, we compare to the other 3-site correlator CGTNS parameterizations. One 
can understand from Table \ref{tab:sextuplet} that the lowest energy, -1542.197777 Hartree, is obtained with the $\Psi_{\rm CGTNS}^{3s}$ 
ansatz (the error is -11.843 mHartree), while the least accurate energy, -1542.195283 Hartree, is obtained from the $\Psi_{\rm CGTNS}^{3s[2s]}$ 
ansatz. Note that in the best case the energy is lowered by only -0.003705 Hartree, which is lower by -0.002951 Hartree compared 
to the energy from $\Psi_{\rm CGTNS}^{3s[2s]{\rm sel}}$. And if one compares the energy from $\Psi_{\rm CGTNS}^{3s[2s]{\rm sel}}$ 
to the conceptually similar ansatz $\Psi_{\rm CGTNS}^{3s[2s]}$, the difference is -0.000457 Hartree, which is rather small. Hence, 
although 3-site correlators do not significantly improve on the total electronic energy, the $\Psi_{\rm CGTNS}^{3s[2s]{\rm sel}}$ 
ansatz is a good approximation to $\Psi_{\rm CGTNS}^{3s[2s]}$.

If all {\it si} correlators are omitted, the $\Psi_{\rm CGTNS}^{3s/si}$ ansatz will minimize the energy to -1542.195739 Hartree, 
while $\Psi_{\rm CGTNS}^{3s/si[2s]}$ yields a slightly higher energy of -1542.195415 Hartree. In both cases, the hybrid 
ans\"atze give higher energies than those from pure 3-site correlator schemes. We note that $\Psi_{\rm CGTNS}^{3s/si[2s]}$ 
is somewhat lower in energy than $\Psi_{\rm CGTNS}^{3s[2s]}$, while $\Psi_{\rm CGTNS}^{3s}$ is far lower than $\Psi_{\rm CGTNS}^{3s/si}$. 
The Monte Carlo parallel tempering optimization procedure fails to decrease the error by more than -0.01 Hartree, even though the 
3-site correlator schemes have much more variational parameters than CAS-SCF. Obviously, the CGTNS ansatz introduces a highly non-linear 
parameterization which requires non-trivial optimization techniques to avoid local minima. Various temperature sets were used for the 
Monte Carlo parallel tempering optimization and only best results are reported in this work. The convergence behavior for all 
parameterizations is shown in Figure \ref{fig:conv_sext3}. One of the problems are rare swap moves between neighboring temperatures. 
If the temperature set has a smaller step, many swaps occur, but this is equivalent to performing simulations on similar sets. For 
the hybrid parameterizations the parallel tempering scheme does not work well because the energies resulting from every temperature set 
are biased to the already optimized $\Psi_{\rm CGTNS}^{2s}$ correlators. This, in turn, increases $\Delta E$ between the neighboring 
temperatures and the swap moves are not likely to appear, see Eq.~(\ref{eq:swap_prob}). This can be easily seen from the convergence 
curves in Figure \ref{fig:conv_sext3}, where we observe steep steps for $\Psi_{\rm CGTNS}^{3s}$ and $\Psi_{\rm CGTNS}^{3s/si}$, 
while all the hybrid parameterizations show a smooth convergence behavior. A solution might be to use dynamically optimized 
temperatures\cite{katz2006}. However, the accuracy achieved here is sufficient for the analysis of CGTNS parameterizations. We emphasize
that accurate relative energies are the ultimate target for processes in low energy chemical physics.

\begin{figure}[H]
\caption{Convergence behavior of hybrid CGTNS parameterizations for manganocene in the lowest-energy sextet state.
\label{fig:hybrid_sext}}
\begin{center}
\includegraphics[scale=0.6]{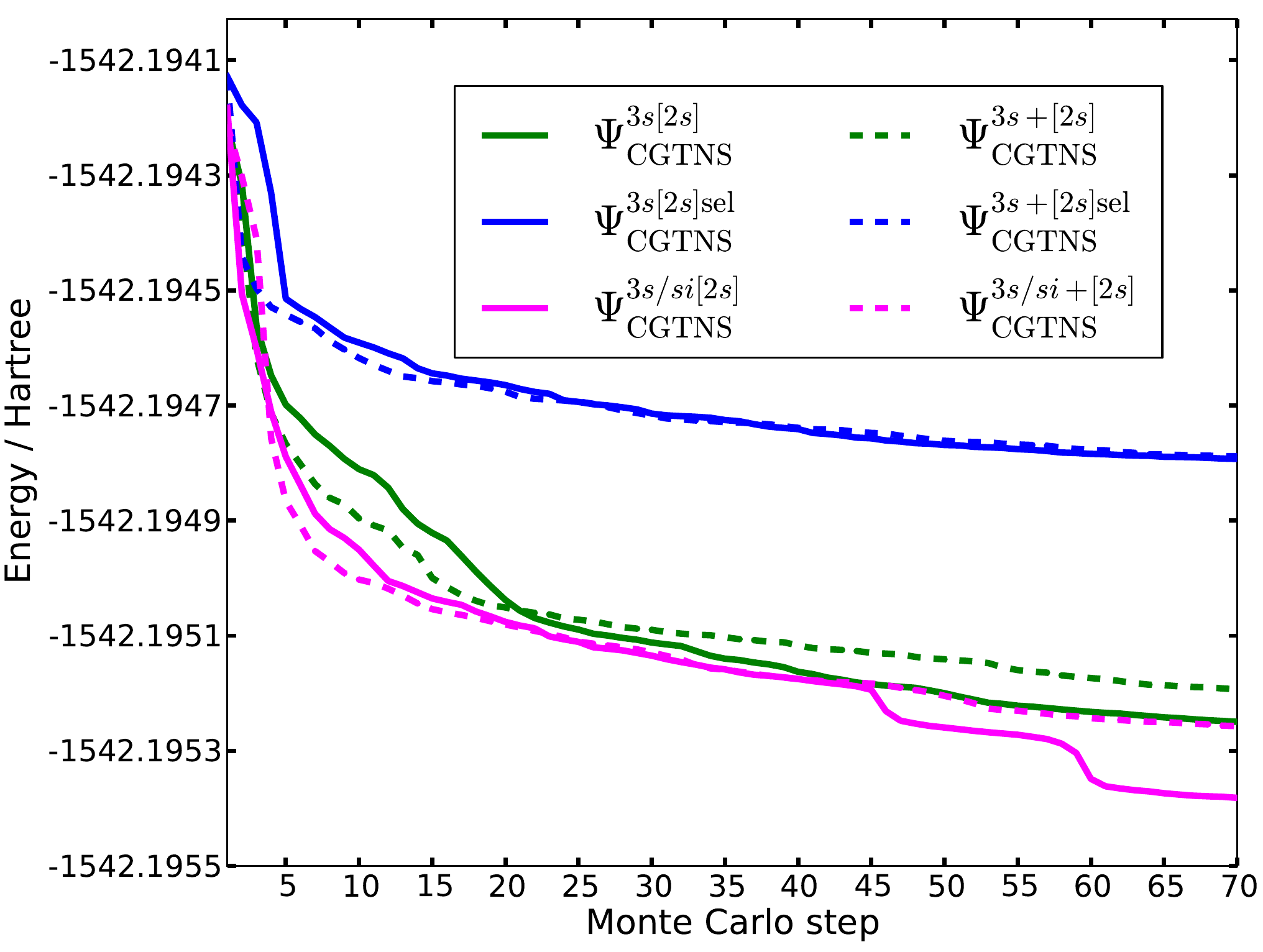}	
\end{center}
\end{figure}

In addition, calculations employing alternative $\Psi_{\rm CGTNS}^{3s+[2s]}$, $\Psi_{\rm CGTNS}^{3s/si+[2s]}$, and $\Psi_{\rm CGTNS}^{3s+[2s]{\rm sel}}$ 
hybrid parameterizations were performed. One can see from Table \ref{tab:sextuplet} that the final energies calculated from $\Psi_{\rm CGTNS}^{3s+[2s]}$, 
$\Psi_{\rm CGTNS}^{3s/si+[2s]}$, and $\Psi_{\rm CGTNS}^{3s+[2s]{\rm sel}}$ wave functions are slightly higher than those of the corresponding $\Psi_{\rm CGTNS}^{3s[2s]}$, 
$\Psi_{\rm CGTNS}^{3s/si[2s]}$, and $\Psi_{\rm CGTNS}^{3s[2s]{\rm sel}}$ parameterizations, respectively. In the case of $\Psi_{\rm CGTNS}^{3s/si+[2s]}$, 
the energy is higher than the one from $\Psi_{\rm CGTNS}^{3s/si[2s]}$ by only 0.125 mHartree, while in the other cases the difference 
is even smaller. The only advantage of the alternative hybrid schemes is slightly faster convergence which can be seen in Figure \ref{fig:hybrid_sext}. 

If the (local) gradient optimization, Eq.~(\ref{eq:grad}), is applied after Monte Carlo optimization, the energy is decreased by only 
-0.34909 mHartree to -1542.19442146 Hartree. With the ``reduced'' gradient optimization, Eq.~(\ref{eq:grad_site}), the 
result deteriorates. The same holds true for the variational optimization suggested by Changlani {\it et al.} in their work on CPS 
ansatz\cite{chan2009} when applied for CGTNS ansatz. Apparently, while such a variational optimization works well in the case of CPS 
ansatz\cite{chan2009}, the presence of correlators different from the ones corresponding to the nearest neighbors introduces 
unsurmountable difficulties. 

\subsection{Manganocene --- Doublet State}\label{sec:doublet}
\begin{figure}[H]
\caption{Convergence behavior of 3-site correlator CGTNS parameterizations for manganocene in the doublet state.
\label{fig:conv_doub}}
\begin{center}
\includegraphics[scale=0.6]{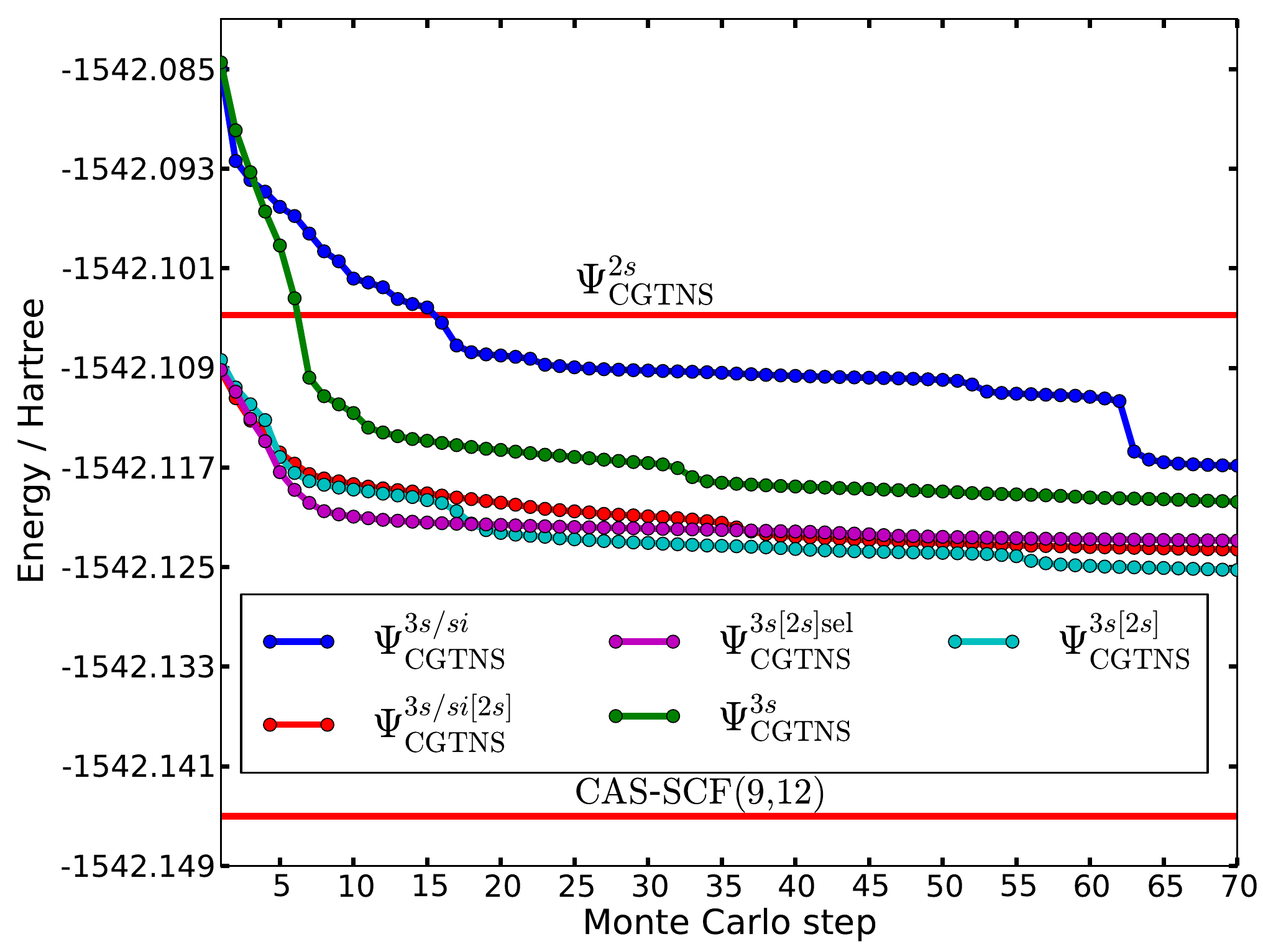}	
\end{center}
\end{figure}

The configurational space of the doublet state is spanned by 98060 ONVs in $C_{2v}$ point group symmetry for an active space consisting 
of 9 electrons in 12 orbitals, see Figure~\ref{fig:orbitals}. But in contrast to the sextet case, the number of spin-adapted CSFs is 
47240 and therefore about half as large as the number of ONVs. It should be emphasized that the number of variational parameters for 
the CGTNS ansatz here is the same as in the case of the sextet, as it depends only on the number of spin orbitals in the active space. 
The 2-site correlator CGTNS scheme reduces the number of variational parameters by 99\% with an error of about 40 mHartree, see 
Table \ref{tab:doublet2}. $\Psi_{\rm CGTNS}^{3s}$ introduces 20800 variational parameters, which corresponds to a 80\% reduction. 
For $\Psi_{\rm CGTNS}^{3s/si}$ the reduction is larger (85\%), corresponding to 16192 variational parameters. 
\begin{table}[H]
\caption{Electronic energies for the doublet state of manganocene calculated with various CGTNS parameterizations and CAS-SCF(9,12).
A percentage indicates a decreased parameter space compared to the 98060 CI coefficients in the CAS-SCF(9,12) reference calculation.}
\label{tab:doublet2}
\begin{center}
\begin{tabular}{lrrr}
\hline\hline
parameterization & parameters & percentage & energy/Hartree\\
\hline
 CAS-SCF(9,12) 			        & 98060 &       & $-1542.144937$\\
 $\Psi_{\rm CGTNS}^{2s}$                & 1200  & 99\% & $-1542.104681$\\
 $\Psi_{\rm CGTNS}^{3s/si}$             & 16192 & 85\% & $-1542.116784$\\ 
 $\Psi_{\rm CGTNS}^{3s}$                & 20800 & 80\% & $-1542.119695$\\
 $\Psi_{\rm CGTNS}^{3s/si[2s]}$         & 16192 & 85\% & $-1542.123527$\\ 
 $\Psi_{\rm CGTNS}^{3s[2s]}$            & 20800 & 80\% & $-1542.125171$\\
 $\Psi_{\rm CGTNS}^{3s[2s]{\rm sel}}$   & 9120  & 91\% & $-1542.122804$\\ 
 $\Psi_{\rm CGTNS}^{3s+[2s]{\rm sel}}$  & 9120  & 91\% & $-1542.123885$\\
\hline\hline
\end{tabular}
\end{center}
\end{table}
The $\Psi_{\rm CGTNS}^{3s/si}$ and $\Psi_{\rm CGTNS}^{3s}$ parameterizations decrease the energy by -0.012103 and -0.015014 Hartree, 
respectively, see Table \ref{tab:doublet2}. From Figure \ref{fig:conv_doub}, it is obvious that 3-site correlators are important for 
energy minimization. As in the case of the sextet state, convergence is accelerated with information obtained from the 2-site correlators. 
As expected, incorporating 3-site correlators into the 2-site correlator ansatz ($\Psi_{\rm CGTNS}^{3s/si[2s]}$ and $\Psi_{\rm CGTNS}^{3s[2s]}$) 
yields energies close (and lower) to the one of the 2-site correlator ansatz right from the start, see Figure 
\ref{fig:conv_doub}. In contrast to the sextet state, both hybrid parameterizations, $\Psi_{\rm CGTNS}^{3s/si[2s]}$ and $\Psi_{\rm CGTNS}^{3s[2s]}$, 
minimize the doublet energies further than the pure 3-site correlator schemes, yielding -1542.123527 and -1542.125171 Hartree, 
respectively. 

As for the sextet, we study whether all of the 3-site correlators are equally important in the wave function parameterization. Again, we 
choose the 18 most entangled spin orbitals according to the selection criterion in Section~\ref{sec:comp_det}, resulting from natural orbitals 
\textbf{D-1}, \textbf{D-2}, \textbf{D-4}, \textbf{D-5}, \textbf{D-6},\textbf{D-9}, \textbf{D-10}, \textbf{D-11}, \textbf{D-12}, see 
Figure~\ref{fig:orbitals}. The corresponding hybrid CGTNS ansatz, $\Psi_{\rm CGTNS}^{3s[2s]{\rm sel}}$, has only 9120 variational parameters 
compared to 20800 in the $\Psi_{\rm CGTNS}^{3s[2s]}$ ansatz. We observe that the chosen 3-site correlators are important for energy minimization. 
One can clearly see in Figure \ref{fig:conv_doub} that $\Psi_{\rm CGTNS}^{3s[2s]{\rm sel}}$ shows even faster convergence and the final energy 
is close to those of $\Psi_{\rm CGTNS}^{3s/si[2s]}$ and $\Psi_{\rm CGTNS}^{3s[2s]}$. Hence, the concept to include higher-order correlators in 
the CGTNS ansatz only for the most entangled orbitals is efficient. 

\begin{figure}[H]
\caption{Convergence behavior of hybrid CGTNS parameterizations for manganocene in the doublet state.
\label{fig:hybrid_doub}}
\begin{center}
\includegraphics[scale=0.6]{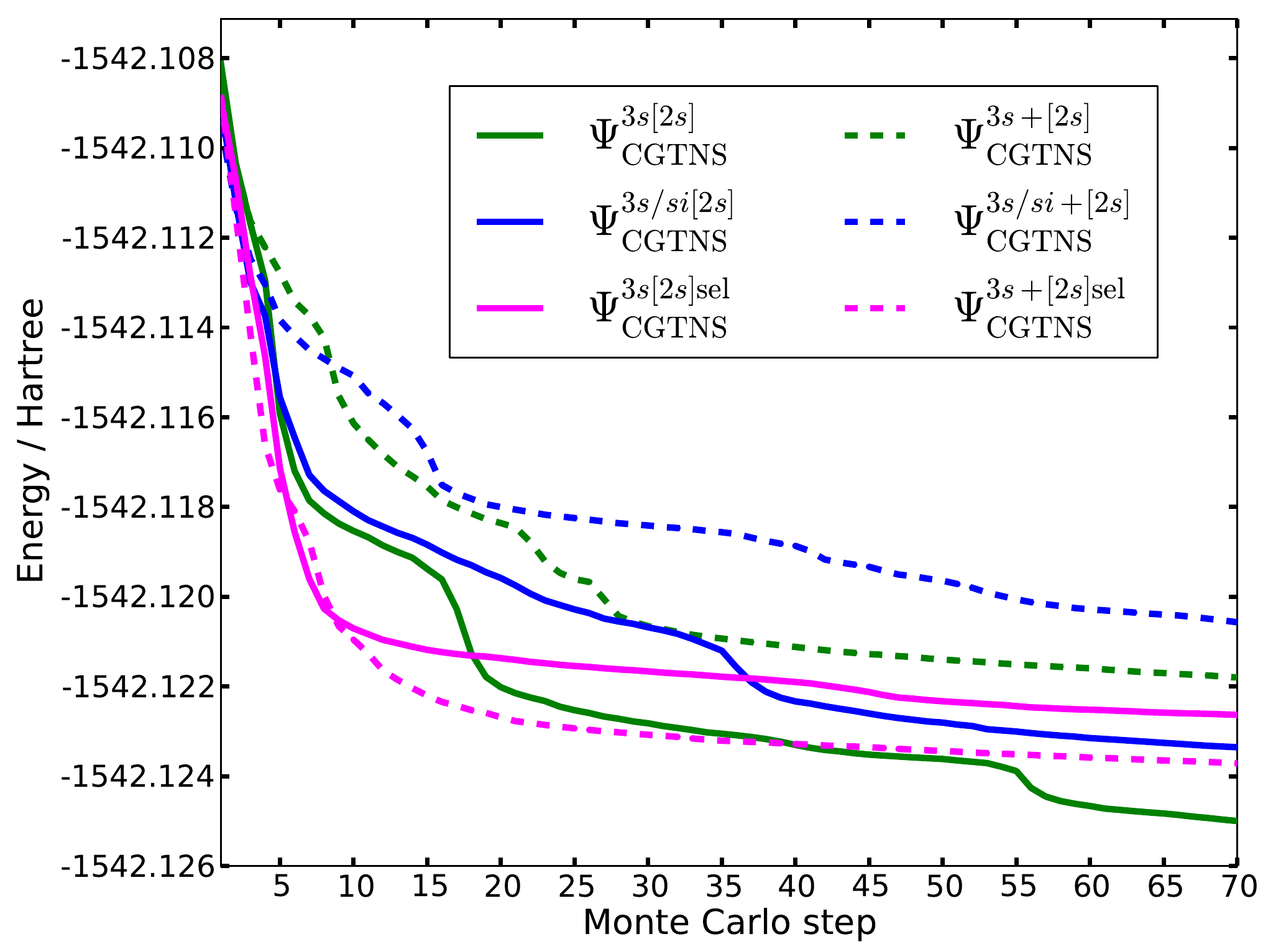}	
\end{center}
\end{figure}

In contrast to the sextet case, the hybrid schemes that employ a sum of 2-site and 3-site correlator products, $\Psi_{\rm CGTNS}^{3s+[2s]}$ and 
$\Psi_{\rm CGTNS}^{3s/si+[2s]}$, show slower energy convergence than their analogs employing products between 2-site and 3-site correlators,
$\Psi_{\rm CGTNS}^{3s[2s]}$ and $\Psi_{\rm CGTNS}^{3s/si[2s]}$, see Figure~\ref{fig:hybrid_doub}. The final energies obtained from 
$\Psi_{\rm CGTNS}^{3s+[2s]}$ and $\Psi_{\rm CGTNS}^{3s/si+[2s]}$ are higher than the ones obtained from their counterparts. An interesting
behavior is exhibited by the $\Psi_{\rm CGTNS}^{3s+[2s]{\rm sel}}$ ansatz, see Figure~\ref{fig:hybrid_doub}. Although at the beginning, it shows 
convergence similar to that of the $\Psi_{\rm CGTNS}^{3s[2s]{\rm sel}}$ ansatz, at the point where the energy from the $\Psi_{\rm CGTNS}^{3s[2s]{\rm sel}}$ 
ansatz is almost converged, the $\Psi_{\rm CGTNS}^{3s+[2s]{\rm sel}}$ ansatz succeeded to overcome local minima and the energy decreases further by 
1.081 mHartree, see Table~\ref{tab:doublet2} and Figure~\ref{fig:hybrid_doub}. 

\subsection{Accuracy Measure}\label{sec:measure}
So far, we observed two peculiarities of the CGTNS ansatz. In the case of the sextet state, the 2-site correlator parameterization provides an 
accurate approximation to the reference wave function and the introduction of higher-order correlators hardly changes the energy any 
further, in the best case by only -0.003705 Hartree. By contrast, $\Psi_{\rm CGTNS}^{2s}$ does not provide an adequate energy estimate 
for the doublet state of manganocene and 3-site correlators are needed to decrease the energy by -0.02049 Hartree. In the light of these observations, it is 
necessary to introduce a measure for the accuracy of CGTNS parameterizations and for the decision whether higher-order correlators should be introduced 
or not. For example, the small energy difference of $\Psi_{\rm CGTNS}^{3s[2s]}$ compared to its extension to 3-site correlators in the 
$\Psi_{\rm CGTNS}^{3s[2s]}$ ansatz, 
\begin{equation}
\Delta E_{\rm CGTNS}^{2s/3s} = E_{\rm CGTNS}^{3s[2s]}-E_{\rm CGTNS}^{2s},
\end{equation}
clearly shows that the introduction of 3-site correlators will not improve the energy. Such an energy-difference measure can serve for 
accuracy control. In the general case of $n$-site correlators, it is given by
\begin{equation}
\Delta E_{\rm CGTNS}^{n/(n+1)s} = E_{\rm CGTNS}^{(n+1)s[ns]}-E_{\rm CGTNS}^{ns}.
\end{equation}
As the next tier of approximation, $\Psi_{\rm CGTNS}^{(n+1)s[ns]}$, may dramatically inflate the variational space, we insert the 
approximation $\Psi_{\rm CGTNS}^{(n+1)s[ns]{\rm sel}}$ instead of $\Psi_{\rm CGTNS}^{(n+1)s[ns]}$,
\begin{equation}\label{eq:measure}
\Delta E_{\rm CGTNS}^{n/(n+1)s} \approx E_{\rm CGTNS}^{(n+1)s[ns]{\rm sel}}-E_{\rm CGTNS}^{ns}.
\end{equation}
The smaller variational space of a $\Psi_{\rm CGTNS}^{(n+1)s[ns]{\rm sel}}$ parameterization makes optimizations feasible and the energy 
usually converges within comparatively few Monte Carlo steps.

\subsection{Manganocene --- Spin-State Splitting}
From Sections \ref{sec:sextet} and \ref{sec:doublet} it is obvious that energies obtained from low-order CGTNS parameterizations still deviate from the 
CAS-SCF reference. This may only be tolerated if relative energies are obtained with higher accuracy. We investigate this issue now for the spin-state 
splitting in manganocene.  As we neglect dynamical correlation, our reference splitting is -40.59 kcal/mol (Table \ref{tab:ref_energy}). 
Since DFT results scatter by about 
30 kcal/mol (Table \ref{tab:ref_energy}, see also \cite{salo2002}), we may consider a deviation by up to about 5 kcal/mol (10\% of the CAS-SCF reference
result) acceptable. 
If the simple strategy to use the same parameterization for 
both spin states is followed, two issues arise. The first one can be clearly seen if one takes the $\Psi_{\rm CGTNS}^{2s}$ ansatz for both spin states. 
For the sextet state, an accurate total energy is obtained, whereas for the doublet state the reduction of the variational space by 99\% is so large 
that it leads to an inaccurate total energy. The energy difference then mounts to -56.09 kcal/mol, see Table \ref{tab:difference}. The problem here 
is that the configurational space for the doublet is 10 times larger than the one for the sextet, while the number of CGTNS variational parameters is 
equal. The second issue can be observed for the $\Psi_{\rm CGTNS}^{3s[2s]}$ parameterization applied for both spin states. Then, the energy difference 
is equal to -44.00 kcal/mol, whereas the variational space for the sextet was enlarged by 59\% relative to the CAS-SCF configurational space. Using 
the same ansatz for both states therefore introduces an imbalance in the approximation of the two different configurational spaces. 

A reasonable strategy is to choose those parameterizations which reduce the variational space by the same amount and are therefore likely to be affected 
by similar errors. For manganocene, $\Psi_{\rm CGTNS}^{3s[2s]{\rm sel}}$ achieves a reduction by 91\% of the variational space for the doublet state, 
while for the sextet state the same reduction is achieved by the $\Psi_{\rm CGTNS}^{2s}$ ansatz. The energy difference estimated based on these schemes 
is -44.72 kcal/mol. If the $\Psi_{\rm CGTNS}^{3s+[2s]{\rm sel}}$ scheme is used instead, the energy difference is equal to -44.04 kcal/mol.

\begin{table}[H]
\caption{The doublet--sextet energy differences in Hartree and kcal/mol for manganocene calculated with various CGTNS parameterizations and CAS-SCF(9,12).}
\label{tab:difference}
\begin{center}
\begin{tabular}{rrrr}
\hline\hline
\multicolumn{2}{c}{parameterization}				& \multicolumn{2}{c}{$E[^6A_1]-E[^2A_1]$} \\
 $^6A_1$ & $^2A_1$ & Hartree & kcal/mol \\
\hline
CAS-SCF(9,12)		             & CAS-SCF(9,12)	                    & $-0.064683$ & $-40.59$ \\
$\Psi_{\rm CGTNS}^{2s}$	             & $\Psi_{\rm CGTNS}^{2s}$              & $-0.089391$ & $-56.09$ \\
$\Psi_{\rm CGTNS}^{3s/si[2s]}$       & $\Psi_{\rm CGTNS}^{3s/si[2s]}$       & $-0.071888$ & $-45.11$ \\
$\Psi_{\rm CGTNS}^{3s[2s]}$          & $\Psi_{\rm CGTNS}^{3s[2s]}$          & $-0.070112$ & $-44.00$ \\
$\Psi_{\rm CGTNS}^{3s/si}$           & $\Psi_{\rm CGTNS}^{3s/si}$           & $-0.078955$ & $-49.55$ \\
$\Psi_{\rm CGTNS}^{3s}$              & $\Psi_{\rm CGTNS}^{3s}$              & $-0.078082$ & $-49.00$ \\
$\Psi_{\rm CGTNS}^{3s[2s]{\rm sel}}$ & $\Psi_{\rm CGTNS}^{3s[2s]{\rm sel}}$ & $-0.072022$ & $-45.20$ \\
$\Psi_{\rm CGTNS}^{3s+[2s]{\rm sel}}$ & $\Psi_{\rm CGTNS}^{3s+[2s]{\rm sel}}$ & $-0.070936$ & $-44.51$ \\
$\Psi_{\rm CGTNS}^{2s}$              & $\Psi_{\rm CGTNS}^{3s[2s]{\rm sel}}$ & $-0.071269$ & $-44.72$ \\
$\Psi_{\rm CGTNS}^{2s}$              & $\Psi_{\rm CGTNS}^{3s+[2s]{\rm sel}}$ & $-0.070187$ & $-44.04$ \\
\hline\hline
\end{tabular}
\end{center}
\end{table}

The energy differences calculated with other 3-site correlator schemes are presented in Table \ref{tab:difference}. The $\Psi_{\rm CGTNS}^{3s/si[2s]}$ 
ansatz performs slightly worse than the $\Psi_{\rm CGTNS}^{3s[2s]}$ ansatz for the spin-state splitting energy giving -45.11 kcal/mol. Approximately 
the same result, -45.20 kcal/mol, can be obtained if the $\Psi_{\rm CGTNS}^{3s[2s]{\rm sel}}$ ansatz is employed in both spin-state calculations. The 
energy difference obtained from the $\Psi_{\rm CGTNS}^{3s+[2s]{\rm sel}}$ ansatz is equal to -44.51 kcal/mol. Whereas in the case of the 3-site CGTNS 
schemes employing correlators for all spin orbitals the variational space for the sextet state is enlarged compared to the CAS-CI space in a reference 
calculation, the $\Psi_{\rm CGTNS}^{3s[2s]{\rm sel}}$ and $\Psi_{\rm CGTNS}^{3s+[2s]{\rm sel}}$ ans\"atze reduce the variational space not only for the 
doublet state but also for the sextet state (66\%). The results obtained from $\Psi_{\rm CGTNS}^{3s/si}$ and $\Psi_{\rm CGTNS}^{3s}$ parameterizations 
are nearly the same and equal to -49.55 kcal/mol and -49.00 kcal/mol, respectively. Note that these schemes enlarge the variational space for the 
sextet to the same extent as their hybrid analogs.

\section{Conclusions}
In this paper, we presented a rigorous analysis of various $n$-site correlators schemes for tensor network states at the example of manganocene.
We demonstrated that the 2-site correlator CGTNS scheme achieves an efficient parameter reduction for the systems with a configurational space 
spanned by about 15 000 ONVs. In the case of the sextet state of manganocene, the number of variational parameters is reduced by 91\% without 
significant loss of accuracy; the error is about 15.5 mHartree. 

Introducing higher-order correlators increases the accuracy and delivers results closer to the CAS-SCF reference results. This, however, comes 
with an unreasonable inflation of the parameter space. We suggested and analyzed two different strategies for the introduction of higher-order 
correlators. The first one assumes that the ansatz features only 3-site correlators and optimizes the energy with respect to them alone. 
The second strategy incorporates 3-site correlators into a converged 2-site correlator ansatz. We demonstrated that such a hybrid extension of 
a 2-site correlator ansatz converges better and faster than the optimization of only 3-site correlators. The 3-site correlator CGTNS parameterizations 
are accurate for the description of a configurational space of about 100 000 ONVs. One can avoid inflation of the variational space by 
considering 3-site correlators only for the most entangled spin orbitals. Such a restriction has only a slight affect on accuracy with respect 
to the original ansatz; in addition, it also increases energy  convergence in the Monte Carlo optimization.

Our results for the doublet state of manganocene showed that a reduction of the variational space by 80\%, 85\%, 91\%, and 99\% leads to errors 
of only -19.766 mHartree, -21.41 mHartree, 23.174 mHartree, and -40.256 mHartree, respectively. Therefore, only an adaptive CGTNS ansatz is promising
that introduces higher-order correlators selectively on demand. 
With the energy measures introduced for accuracy control such an ansatz can adapt to the electronic structure under the study.

A reduction of variational space by 85\% to 90\% leads to a 20 mHartree error for total electronic energies. However, for most chemical processes, 
the evaluation of energy {\it differences} is most important. We found that reliable results can be obtained if the same reduction of variational parameters 
for the two energies to be compared is achieved. For the manganocene sextet--doublet energy difference, the error introduced by tensor network parameterizations can be 
reduced to only 8.5\% of the reference value. 

While the non-stochastic optimization schemes performed well for the CPS ansatz, for CGTNS they turned out to be inefficient. Hence, 
additional work on the improvement of our Monte Carlo optimization scheme is required. A possibility is the introduction of dynamically optimized temperature 
sets\cite{katz2006}.

\section{Acknowledgment}
This work was supported by ETH Zurich (ETH Fellowship FEL-27 14-1).


\end{document}